\documentclass[draft]{agujournal2019}
\usepackage{url}
\usepackage{soul}
\usepackage[latin1]{inputenc}
\usepackage{graphicx}
\usepackage{tikz}
\usetikzlibrary{quotes,arrows.meta,shapes,positioning,shapes.geometric}

\usepackage{pgfplots}
\usepackage{subcaption}
\usepackage{amsmath}
\usepackage{amsfonts}
\usepackage{siunitx}
\usepackage{natbib}

\usepackage{caption}
\usepackage{subcaption}

\newcommand{\changed}[1]{#1}
\newcommand{\changetwo}[1]{#1}

\draftfalse

\begin{document}

\pagestyle{plain} 

\title{Bonded discrete element simulations of sea ice with non-local failure: Applications to Nares Strait}

\authors{Brendan West\affil{1}, Devin O'Connor\affil{1}, Matthew Parno\affil{2}, Max Krackow\affil{1}, Chris Polashenski\affil{1,3}}

\affiliation{1}{Cold Regions Research and Engineering Laboratory, U.S. Army Corps of Engineers, Hanover, NH, USA}
\affiliation{2}{Department of Mathematics, Dartmouth College, Hanover, NH, USA}
\affiliation{3}{Thayer School of Engineering, Dartmouth College, Hanover, NH, USA}

\correspondingauthor{B. West}{brendan.a.west@erdc.dren.mil}

\begin{keypoints}
	\item The DEM with bonded particles and physics-based fracture models can \changed{qualitatively capture the behavior of sea ice flowing through a channel.}
	\item Fracture is captured with a non-local stress calculation and Mohr-Coulumb failure model to determine when inter-particle bonds fail.
	\item \changed{We use spatio-temporal scaling analyses to quantitatively assess the model's ability to capture key properties of sea ice deformation.}
\end{keypoints}

\begin{abstract}
\changed{The discrete element method (DEM)} can provide detailed descriptions of sea ice dynamics that explicitly model \changed{floes and} discontinuities in the ice, which \changed{can be challenging to represent accurately with current models. However, floe-scale stresses that inform lead formation in sea ice are difficult to calculate in current DEM implementations. In this paper, we use the ParticLS software library to develop a DEM that models} the sea ice as a collection of discrete rigid particles that are initially bonded together using a cohesive beam model that approximates the response of an Euler-Bernoulli beam located between particle centroids. Ice fracture and lead formation are determined based on the value of a non-local \changed{Cauchy} stress state around each particle and a Mohr-Coulomb fracture model. Therefore, large ice floes are modeled as continuous objects made up of many bonded particles that can interact with each other, deform, and fracture. We generate particle configurations by discretizing the ice in MODIS satellite imagery into polygonal floes that fill the \changed{observed ice shape and extent}. The model is tested on ice advecting through an idealized channel and through Nares Strait. \changed{The results indicate that the bonded DEM model is capable of qualitatively capturing the dynamic sea ice patterns through constrictions such as ice bridges, arch kinematic features, and lead formation. In addition, we apply spatial and temporal scaling analyses to illustrate the model's ability to capture heterogeneity and intermittency in the simulated ice deformation.}
\end{abstract}

\section*{Plain Language Summary}
\changed{Numerical} models of sea ice give researchers important tools to study how the Arctic is changing. Discrete element method (DEM) models \changed{idealize sea} ice as a collection of individual rigid bodies\changed{, or ``particles,''} that can interact with each other independently, and can capture the discontinuities and geometric force concentrations in ice that are common at small scales. In this paper, we extend \changed{recent DEM models and evaluate a non-local} stress state within the modeled ice (bonded DEM particles) to determine when the ice should fracture. As a result, the model simulates large pieces of ice that can break into smaller pieces, or floes, composed of many still-bonded particles. This allows us to represent both discrete fractures, and emergent aggregate behavior of ice as it deforms. As an example, we simulate ice advecting through Nares Strait. 
\sloppy

\section{Introduction}

\changed{Numerical models of sea ice play an important role in understanding the changing Arctic and allow researchers to predict the dynamic response of sea ice to different environmental conditions. High resolution forecasts from predictive models are also becoming increasingly important due to increased human activity in the Arctic.} The recent decline in Arctic sea ice has lead to more traffic in the Arctic Ocean for fishing, resource extraction, tourism, cargo shipping, and military purposes. \changed{Sea ice models} that can explicitly capture small discontinuities and fractures in the ice are particularly valuable for navigation. For example, \citet{Mariner2019} lists high resolution information about compression and pressure ridges as one of the most important things missing in current operational ice products.

Many sea ice models, such as those used in global climate models, employ continuum approaches where the sea ice is discretized with an Eulerian mesh and the ice is modeled with constitutive models such as viscous-plastic (VP) or elastic-viscous-plastic (EVP) rheologies \citep{Hibler1979, Hunke1997}. \changetwo{Recent studies, such as \citep{Bouchat2017} and \citep{Hutter2020}, have shown that VP/EVP rheologies can capture important statistics about largescale sea ice deformation. On smaller scales however,} it has been shown that the VP rheologies can be inconsistent with observed stress and strain-rate relationships \citep{Weiss2007}, tensile strength \citep{Coon2007}, ridge distribution \citep{schulson2004compressive}, and lead intersection angles \citep{ringeisen2019simulating}. Efforts to overcome the limitations of VP rheologies are typically either focused on the development of new rheologies (e.g., \changed{\citet{Schreyer2006, Wilchinsky2006, Girard2011, Dansereau2016})} or on the development of discrete techniques, like the discrete element method (DEM), that adopt a Lagrangian viewpoint and \changed{model} the interaction of individual ice particles. \changed{Other novel methods include the  material point method \citep{Sulsky2007} which blurs the lines between an Eulerian and pure Langrangian model, or the neXtSIM finite element model \citep{Rampal2016} that takes a Langragian perspective with adaptive re-meshing.}

\changed{Several} efforts have used the DEM to simulate sea ice dynamics  \citep{Hopkins2004,Hopkins2006,Herman2013a,Herman2016,Kulchitsky2017,Damsgaard2018}. The DEM explicitly models the dynamics of individual rigid bodies, or ``particles'', and can therefore capture discontinuities in sea ice such as cracks and leads that are common near the ice edge or in the marginal-ice-zone (MIZ). The DEM is a promising modeling approach for sea ice \changed{forecasting applications} \citep{Hunke2020}, however many DEM sea ice studies to date have used simplified contact models and particle geometries in order to lessen the computationally-intensive process of tracking and calculating the interaction between many particles. For example, it is common to use elastic, viscous-elastic, or Hertzian contact models to calculate inter-particle forces that do not account for the energy lost due to ridging between ice floes \changed{\citep{Sun2012,Herman2013a,Herman2013b,Herman2016,Kulchitsky2017}}. It is also common to represent particles with disks or simple shapes due to the ease of solving contact between basic shapes \citep{Sun2012, Herman2013a,Herman2016,Damsgaard2018,Jou2019}. Although these modifications increase the speed of the models, oversimplifying the complex geometries and interactions found in real sea ice can limit the \changed{accuracy} of these models.  It has been shown that particle shape greatly affects the bulk behavior of \changed{simulated granular materials} \citep{Kawamoto2016,Kawamoto2018}. In particular, using disk-shaped particles reduces the bulk shear strength of the material as compared to using irregular particle geometries \citep{Damsgaard2018}.

\changed{In this paper we build upon recent recent advances in DEM models and develop a 2D model} that uses cohesively-bonded polygonal-shaped particles, and a non-local physics-based fracture model to capture the behavior of sea ice. Recently, \changed{\citet{Damsgaard2018} presented a simplified DEM model of ice jamming within constrictions, with the goal of developing a computationally efficient DEM model that could be used in global climate models. Although they were able to simulate jamming behavior, they note that the simplified model misses certain aspects of observed sea ice behavior, in part due to their spherical particle shapes and particle contact laws.} We use a new DEM software library called ParticLS \citep{Davis2021} that can represent sea ice floes with convex polygons to better capture the irregular shapes often observed in sea ice. ParticLS implements the cohesive beam model \citep{Andre2012}, which was developed to simulate continuous materials as collections of bonded DEM particles. This cohesive model uses the analytical response of Euler-Bernoulli beams placed between centroids of adjacent particles to propagate stresses and strains through the bonded particle collection. These beams can break, thereby simulating discontinuities in the material. 

Many DEM sea ice models have simulated cohesion between particles, however they have typically evaluated the local stress state within each bond to determine if they should break. \citet{Damsgaard2018} and \citet{Herman2016} compared the maximum normal and maximum shear stresses within the bonds against prescribed thresholds\changed{, whereas \citet{Hopkins2004} decreased the bond stress after a compressive or tensile threshold was reached, thereby gradually weakening the ice post-failure. \citet{Wilchinsky2010} found that bond failure models that only consider tensile and compressive failure can result in unnatural rectilinear crack paths. Therefore, they compared the stresses within each bond against a Mohr-Coulomb failure envelope. A similar approach was used in \citep{Kulchitsky2017}.} We also employ a Mohr-Coulomb failure model due to its well-known ability to describe sea ice fracture, but we extend the approach by evaluating the non-local stress states of each particle to determine whether bonds should fail. This non-local stress approach\changed{, which is similar to \citet{Andre2013},} considers the stress-state produced by all DEM particles within a small neighborhood, which has been shown to reproduce more accurate crack patterns in elastic brittle materials than localized bond fracture models \citep{Andre2013,Andre2017}. We are unaware of applications of either the cohesive beam law or non-local stress evaluations in DEM models of sea ice, or evaluations of their ability to capture salient sea ice behavior.

\begin{figure}[h!]
\begin{center}
\begin{tikzpicture}
    \node[anchor=south west,inner sep=0] (image) at (0,0) {\includegraphics[width=0.56\textwidth]{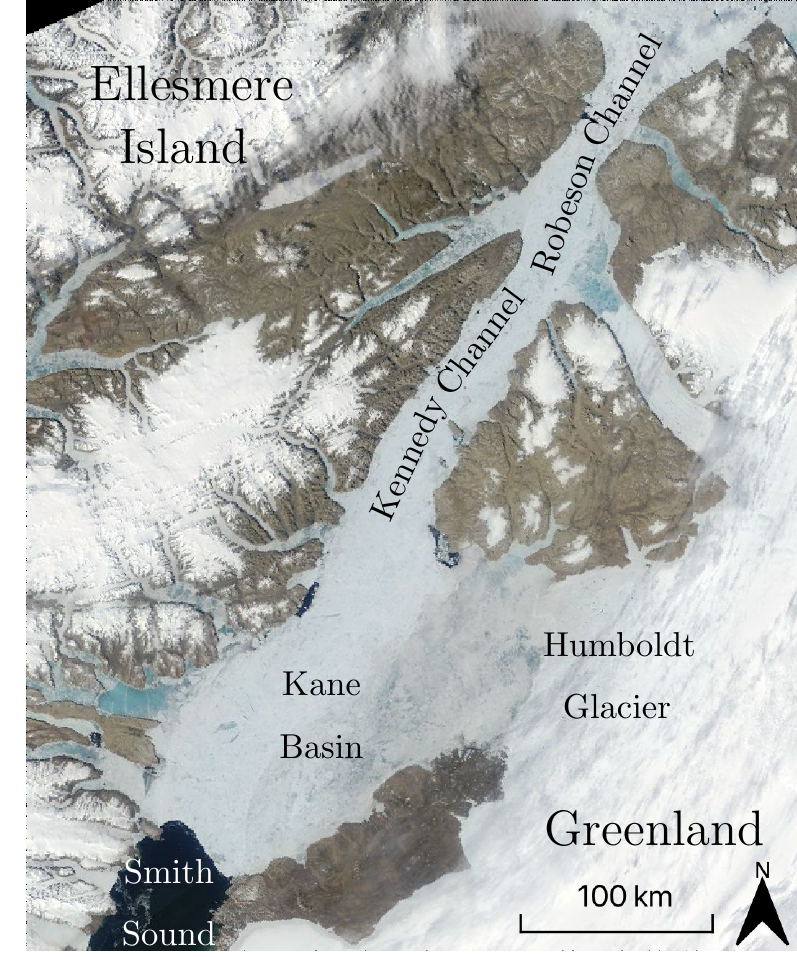}};
\end{tikzpicture}
\end{center}
\caption{Map of Nares Strait region and sub-regions. The underlying MODIS image is from June 28, 2003, and reflects the ice extent and arch from which we initialized the floe DEM collection.}
\label{fig:NaresStraitDomain}
\end{figure}

To test our model, we follow the precedent set by earlier works \citep{Dumont2009,Rasmussen2010,Dansereau2017,Damsgaard2018}, and simulate sea ice advecting through channel domains that encourage arch formation and failure. Ice arches are examples of large-scale sea ice behavior that result from small-scale interactions of ice parcels that jam in constricted regions. The arches form as distinct cracks across the constriction that completely stop and separate the ice upstream from the ice flowing downstream. These arches often result in long-lasting discontinuities in the ice. We use an idealized channel case \changed{from \citet{Dumont2009} and \citet{Dansereau2017} to examine the arching and break up process using our bonded-DEM model. Next, we examine the ice dynamics and arch behavior through Nares Strait (Figure \ref{fig:NaresStraitDomain}). Additionally, we examine the export of ice mass through the strait and explore simulated floe size distributions, both as a function of ice strength. The Nares Strait arches are well-studied features that form within the strait itself, and at the entrance from the Lincoln sea. These arches play important roles in limiting the amount of sea ice flux through that region, but break up almost every spring, resulting in highly-discontinuous sea ice that advects out of the strait~\citep{Moore2021}.} 

In the following sections we describe the governing equations, contact laws, and forcing functions that comprise our model. Section \ref{sec:modeling-overview} describes the momentum balance driving the ice motion, as well as the DEM approach and different models we use to simulate the resultant dynamics. In section \ref{sec:particle-init} we describe the method used to initialize the particles from MODIS imagery. \changed{In Section \ref{sec:scaling_approach}, we present an approach for the spatio-temporal scaling analysis of DEM simulations, which allows us to quantitatively assess our model's ability to capture the heterogeneous and intermittent deformation of sea ice.} Sections \ref{sec:idealized_channel} and \ref{sec:nares-strait} present the results of the idealized channel and Nares Strait simulations, and compares the Nares Strait results with behavior seen in optical satellite imagery. Section \ref{sec:conclusion} discusses the effectiveness of this method in capturing the sea ice dynamics as well as future developments.

\section{Model Overview}
\label{sec:modeling-overview}
The principal forces acting on sea ice include drag from wind and ocean currents ($F_{a}$ and $F_{o}$), internal stress gradients within the ice ($F_{s}$), Coriolis forces ($F_{c}$), and forces due to sea surface tilt ($F_{t}$) \citep{Hibler1979, Steele1997}: 
\begin{linenomath*}
\begin{equation}
    \rho h \frac{du}{dt} = F_{a} + F_{o} + F_{s} + F_{c} + F_{t}
\label{eq:momentum-balance-full}
\end{equation}
\end{linenomath*}
where $\rho$ is ice density, $h$ is ice thickness, and $\frac{du}{dt}$ is the ice acceleration. This force balance generally consists of wind driven forces trying to move the ice, with ocean drag and the internal ice stress resisting the motion \citep{Thorndike1982}. As a result, the motion of ice in free drift is typically dominated by wind and ocean currents, whereas the internal ice stress dominates when the ice is consolidated or constricted \citep{Steele1997}. The Coriolis and surface tilt terms are usually small \citep{Steele1997}, especially for ice dynamics over the span of a few days and over smaller spatial scales \citep{Wadhams2000}. In addition, \citet{Rallabandi2017} notes that the Coriolis force is diminished within narrow straits because the force typically acts normal to the direction of flow. We assume a stagnant ocean current \changed{and constant surface height}. Therefore, we ignore the affects of Coriolis and surface tilt forces acting on the ice in our simulations. In the following sections we describe how the DEM models these forces, including the cohesion model used to capture the internal stress state within consolidated ice and the drag model used to account for wind and ocean forces.

The DEM was first applied to sea ice in the 1990s \citep{Hopkins1991,Loset1994_Pt1,Loset1994_Pt2,Jirasek1995,Hopkins1996}, and it was shown to be an effective method for modeling the interactions between individual ice floes. The DEM discretizes the ice into particles and then uses the balance of linear and angular momentum to define a system of differential equations describing the motion of each particle. The conservation of linear momentum results in
\begin{linenomath*}
\begin{equation}
	m_i \dot{u}_i(t) = \sum_{j=1}^{n} f_{i,j}(t) + f_{i,s}(t),
\label{eq:dem-newton-second-one}
\end{equation}
\end{linenomath*}
where
\begin{itemize}
	\item $m_i$ is the mass of the $\mathit{i\mbox{-}th}$ particle, 
	\item $\dot{u}_i(t)$ is the particle's acceleration,
	\item $f_{i,j}(t)$ is the force acting on particle $i$ from particle $j$,
	\item $f_{i,s}(t)$ are body forces acting on the surfaces of the particle (e.g., drag), 
\end{itemize}
Similarly, the conservation of angular momentum results in 
\begin{linenomath*}
\begin{equation}
	I_i \dot{\omega}_i(t) = \sum_{j=1}^{n} \tau_{i,j}(t) +  \tau_{i,s}(t),
\label{eq:dem-conserve-angular-one}
\end{equation}
\end{linenomath*}
where
\begin{itemize}
	\item $I_{i}$ is the particle's moment of inertia tensor about it's center of mass,
	\item $\dot{\omega}_i(t)$ is the particle's angular acceleration, 
	\item $\tau_{i,j}(t)$ is the torque acting on particle $i$ from particle $j$, 
	\item $\tau_{i,s}(t)$ is the torque from surface forces. 
\end{itemize}
The system of differential equations \eqref{eq:dem-newton-second-one}--\eqref{eq:dem-conserve-angular-one} can then be integrated numerically to evolve the particle positions and orientations. We direct the reader to \citep{Davis2021} for additional information regarding the specifics of the numerical methods used in ParticLS. 

The surface forces, $f_{i,s}$, acting on the particles correspond to drag loads that drive ice particle motion. The inter-particle forces, $f_{i,j}$, and torques, $\tau_{i,j}$, on the other hand, are calculated following a prescribed ``contact law" that describes the material response to these forcings. The contact law depends on properties of the ice pack; a different contact law is required to model ice in free drift compared to pack ice where ice floes are bonded to each other. Below, Section \ref{sec:bonded_law} describes our approach for modeling cohesively bonded particles while Section \ref{sec:ridging_law} describes our approach for modeling non-bonded contact. In Section \ref{sec:failure_model}, we describe a non-local failure criteria, which governs the transition from bonded to non-bonded contact. We believe our approaches for bonded contact and failure are unique in DEM simulations of sea ice. Note that in our simulations, all particles are initially bonded together.

\subsection{Cohesive Contact Law}
\label{sec:bonded_law}
Ice floes are pieces of ice that move as a single cohesive body, whose size and shape change frequently due to fracture and re-freezing. A common approach in DEM models of sea ice is to represent each floe as a particle in the simulation \citep{Hopkins1996,Hopkins2004,Herman2013a,Damsgaard2018}. However, this makes the floes non-deformable. \citet{Hopkins2006} introduced representations of floes as collections of small particles bonded together that can deform via inter-particle bonds. In that work, a viscous-elastic ``glue'' was used to capture tensile and compressive forces between particles. \citet{Herman2016} also simulated floes with multiple bonded particles, however they used disk particles, which inherently leave gaps in the floe. Similar to \citet{Hopkins2006}, we treat the initial consolidated ice pack as a collection of bonded polygons, where the evolution of floe sizes and shapes results from sequential fracture of the inter-particle bonds. However, we employ a different strategy, based on cohesive beams, for bonding particles.  The cohesive bond model simulates the behavior of an Euler-Bernoulli beam to describe the tensile, compressive, and bending forces generated between adjacent bonded particles. The equations that describe the bonded inter-particle forces and moments can be seen in \citep{Andre2012}. This cohesion is important for our simulations, as it has been found that stable ice arches require cohesive strength between individual ice parcels in order to sustain the stress generated in the arch \citep{Hibler2006, Damsgaard2018}. The cohesive beam model we use has not previously been applied to simulations of sea ice, however it has been used to accurately model brittle elastic materials as collections of bonded DEM particles \citep{Andre2012,Andre2013,Andre2017,Nguyen2019}. To retain numerical stability in our simulations and prevent spurious oscillations in our beam forces we add damping proportional to the relative velocity between the particles bonded by the beam. Similar to other bonded sea ice models (e.g., \citet{Hopkins1994}), the value used was calculated based on a proportion of the critical beam damping, $2\zeta_{b}\sqrt{K_bm_i}$, where $\zeta_{b}$ is the beam damping ratio, \changetwo{and $m_i$ is the ice mass. $K_b$ is the beam stiffness, and is calculated with the ratio $E_bA_b/l_b$, where $E_b$ is the beam modulus, $A_b$ is the beam cross-sectional area, and $l_b$ is the beam length, defined as the distance between bonded particle centroids.} The beam parameters used in these simulations are summarized in Table \ref{table:model-params}.

\subsection{Sea Ice Failure Model}\label{sec:failure_model}
\label{sec:failure-model}
The failure criterion for the inter-particle bonds plays a critical role in our analysis, as it dictates how the initial bonded ice pack fractures into smaller floes. Like \citet{Weiss2007}, \citet{Rampal2016}, \changed{\citet{Wilchinsky2010},} and \citet{Kulchitsky2017}, we employ a Mohr-Coulomb failure criterion that accounts for tensile ($\sigma_{N,t}$) and compressive ($\sigma_{N,c}$) failure. Unlike previous sea ice DEM efforts however, we employ a non-local approach for estimating the stress (see discussion below).  The Mohr-Coulomb failure thresholds are
\begin{eqnarray}
	\sigma_1 &\leq& q \sigma_2 + \sigma_c \label{eq:failure-MohrCoulomb} \\
	\frac{\sigma_1 + \sigma_2}{2} &\geq& \sigma_{N,t} \label{eq:failure-tension}\\ 
	\frac{\sigma_1 + \sigma_2}{2} &\leq& \sigma_{N,c}
\label{eq:failure-compression},
\end{eqnarray}
where tension is positive, compression is negative, and $\sigma_1$ and $\sigma_2$ are the principal stresses. $q$ and $\sigma_c$ are defined following \citet{Rampal2016} and \citet{Weiss2009a}:
\begin{eqnarray}
	q &=& \left[(\mu^2 + 1)^{1/2} + \mu\right]^{2}
\label{eq:failure-q}\\
	\sigma_c &=& \frac{2c}{(\mu^2 + 1)^{1/2} - \mu},
\label{eq:failure-sigma_c}
\end{eqnarray}
where $\mu$ is the internal friction coefficient, and $c$ is the cohesion of the ice. This failure criterion has been shown to capture the mechanics of dense granular materials \citep{Damsgaard2018}, as well as the failure envelope seen in physical measurements of sea ice \citep{Weiss2007}. Similar to \citet{Dansereau2017}, we use a uniform distribution between minimum ($c_{min}$) and maximum ($c_{max}$) cohesion values when initializing our DEM particles to create heterogeneity in the ice strength and resultant failure. 

It is well known that bonded lattice-like DEM approaches require calibration of \changed{bond} parameters in order to simulate realistic macroscopic or effective response and failure properties \citep{Andre2019}. Therefore, we created calibration simulations to determine the appropriate failure model values $\sigma_{N,t}$ and $\sigma_{N,c}$. We studied the uniaxial compression and tension of a 154 by 308 km block of ice composed of approximately 4000 bonded particles. The failure parameters were prescribed such that the specimen failed in tension and compression at the effective stresses found in the literature \citep{Weiss2009a} for ice at geophysical scales. We also used these simulations to determine appropriate values for the beam elastic modulus, $E_{b}$, and Poisson's ratio, $\nu_{b}$, for the cohesive model presented in Section \ref{sec:bonded_law}. The beam parameters were prescribed such that the specimen's effective elastic modulus matched values found in the literature for sea ice. These failure stresses and beam parameters are shown in Table \ref{table:model-params}.

\changed{Several sea ice DEM models have based bond failure on the stress within each individual bond \citep{Hopkins2006,Wilchinsky2010,Herman2016,Kulchitsky2017,Damsgaard2018}. As mentioned above, calibration studies are often required to find realistic failure parameters, however in our testing we found that these per-bond failure models were overly-brittle and created large amounts of fragmentation, where large regions of ice disintegrated into many un-bonded particles. These per-bond failure methods do not consider the behavior of nearby bonds, and do not limit the number of bonds that can fail at a time \citep{Hopkins2006,Wilchinsky2010,Herman2016,Kulchitsky2017,Damsgaard2018}. We adapt an alternative approach from \citet{Andre2013} that computes the stress contributions from all neighboring particles within a small region around a given particle. Compared to the stress in individual bonds, this non-local stress provides a more representative evaluation of the stress state at a particle's location. Following \citet{Nguyen2019}, we calculate each particle's symmetric non-local Cauchy stress tensor using
\begin{linenomath*}
\begin{equation}
	\overline{\overline{\sigma}}_{\Omega} = \frac{1}{2\Omega} \bigg( \sum_{j=1}^{N} \frac{1}{2} ( \mathbf{r}_{i,j} \otimes \mathbf{f}_{i,j} + \mathbf{f}_{i,j} \otimes \mathbf{r}_{i,j} ) \bigg),
\label{eq:non-local-stress}
\end{equation}
\end{linenomath*}
where
\begin{itemize}
	\item $\Omega$ is the volume of particle $i$, 
	\item $N$ is the total number of neighboring bonded particles, 
	\item $\otimes$ is the tensor product between two vectors,
	\item $f_{i,j}$ is the force imposed on particle $i$ from the beam between $i$ and $j$, 
	\item $r_{i,j}$ is the vector between the centroids of particles $i$ and $j$.
\end{itemize}
This tensor is calculated at every time step for each particle $i$ using the $N$ adjacent particles that are still bonded to particle $i$. This stress tensor allows us to compute the principal stresses within the ice and compare them against more traditional failure surfaces used to capture sea ice failure, like the Mohr-Coulomb envelope defined above. 

Once the failure criteria is met, a select portion of the particle's bonds are broken. We find the direction of largest tensile principal stress and then define a plane perpendicular to that vector. All bonds that fall on one side of this plane are then severed, as shown in Figure 6 of \citet{Andre2017}. A comparison of non-local and per-beam failure models in DEM simulations was performed by \citet{Andre2013}. They showed that the per-bond failure model resulted in highly-fragmented damage, whereas the non-local model produced fractures that quantitatively matched the linear, continuous fractures measured in indenter experiments of silica glass \citep{Andre2013}. The results presented below suggest that this type of non-local failure model is also able to reproduce the realistic fracture patterns of sea ice flowing through a constriction.}

\subsection{Ridging Contact Law}\label{sec:ridging_law}
\changed{Once the cohesive bonds have broken between two particles, the particles interact through a contact model that approximates the physics of interacting pieces of ice. Many DEM contact laws have been used in the sea ice DEM field, and some 2D contact models have been developed to} approximate out-of-plane behavior, such as pressure ridging, which is an important mechanism for dissipating stress in the ice pack. For particles in free-drift, we adopt the elastic-viscous-plastic contact model developed by \citet{Hopkins1994,Hopkins1996} to approximate the energy lost due to crushing and ridging between contacting floes. The model accounts for two regimes; one where the generated forces are small enough to maintain elastic contact, and a second where the forces are large enough that plastic deformation occurs. In both regimes, the normal force is a function of the overlap area between contacting polygons, with a viscous component related to how quickly the overlap area changes. The tangential loads are calculated with an elastic contact model that is limited by a Coulomb friction limit. \citet{Hopkins1996} provides more details on this contact model. Similar to the cohesive beam model, we add damping proportional to the relative velocity between the particles undergoing ridging contact to retain numerical stability. Following other bonded sea ice models \citep{Hopkins1994}, the value used was calculated based on a proportion of the critical ridging damping, $2\zeta_{r}\sqrt{K_im_i}$, where $\zeta_{r}$ is the ridging damping ratio, $K_i$ is the sea ice stiffness and $m_i$ is the ice mass. The model parameters used in these simulations are adopted from \citet{Hopkins1996}, and are summarized in Table \ref{table:model-params}.

\subsection{Atmosphere and Ocean Drag}
\label{sec:drag}
Drag forces acting on ice due to wind and ocean currents can be described with the following quadratic laws \citep{Hibler1986,Hopkins2004}:
\begin{eqnarray}
    \vec{F}_{a} &=& \rho_{a}C_{a}A_{i}|\vec{v_{a}}|\big(\vec{v_{a}}\cos\theta_a + \hat{k}\times\vec{v_{a}}\sin\theta_a\big)
    \label{eq:wind-drag-force}\\
    \vec{F}_{o} &=& \rho_{o}C_{o}A_{i}|\vec{v_{o}}-\vec{v_{i}}|\big((\vec{v_{o}}-\vec{v_{i}})\cos\theta_o + \hat{k}\times(\vec{v_{o}}-\vec{v_{i}})\sin\theta_o\big)
    \label{eq:ocean-drag-force}
\end{eqnarray}
where the $_a$, $_o$, and $_i$ subscripts correspond to quantities related to the wind, ocean, and the individual particles, respectively. The $\theta_a$ and $\theta_o$ terms are the wind and ocean turning angles, and $\hat{k}$ is a unit vector oriented in the direction normal to the sea ice plane. Often times the turning angles are assumed to be 0, which is also assumed for these simulations, thereby simplifying equations \eqref{eq:wind-drag-force} and \eqref{eq:ocean-drag-force}. It is also commonly assumed that the relative velocity between the air and ice is dominated by the wind, \changetwo{which is why} equation \eqref{eq:wind-drag-force} only considers the wind velocity. In these 2D simulations we account for the skin drag acting on the horizontal surface of the sea ice due to the wind and ocean, and we adopt values for the drag coefficients that are similar to those commonly used in the literature (see Table \ref{table:model-params}) \citep{Hopkins2004,Martin2010,Gladstone2001}. 

The DEM sea ice literature contains several ways of accounting for the torque generated by drag. Some authors ignore it altogether (see e.g., \citet{Hopkins2004,Martin2010}) while others calculate the torque due to ocean drag, but not atmospheric drag \citep{Herman2016}. In reality, torque can result from the curl of ocean and atmosphere currents. \citet{Damsgaard2018} states however, that it is reasonable to ignore the curl of ocean and atmosphere currents on the scale of individual ice floes. Due to the length scales of our simulations we ignore the torque resulting from curl. However, we apply a resistive moment resulting from the ocean drag, similar to \citet{Hopkins2001}, \citet{Sun2012} and \citet{Herman2016}, but accounting for only the drag on the submerged horizontal surface of the floe:
\begin{linenomath*}
\begin{equation}
	M_{o} = -\rho_{o}r^{3}C_{o,h}A_{o,h}|\omega|\omega,
\label{eq:ocean-drag-torque}
\end{equation}
\end{linenomath*}
where $r$ is the polygonal floe's effective moment arm, and $\omega$ is the floe's angular velocity in the z-direction. We assume the resistive moment due to wind is minimal and therefore ignore it. Due to the 2D nature of these simulations, these moments result in reduced rotation around the z-direction.

\section{Particle Initialization}
\label{sec:particle-init}
\begin{figure}[!htbp]
\begin{center}
\begin{tikzpicture}
    \node[anchor=south west,inner sep=0] (image) at (0,0) {\includegraphics[width=0.9\textwidth]{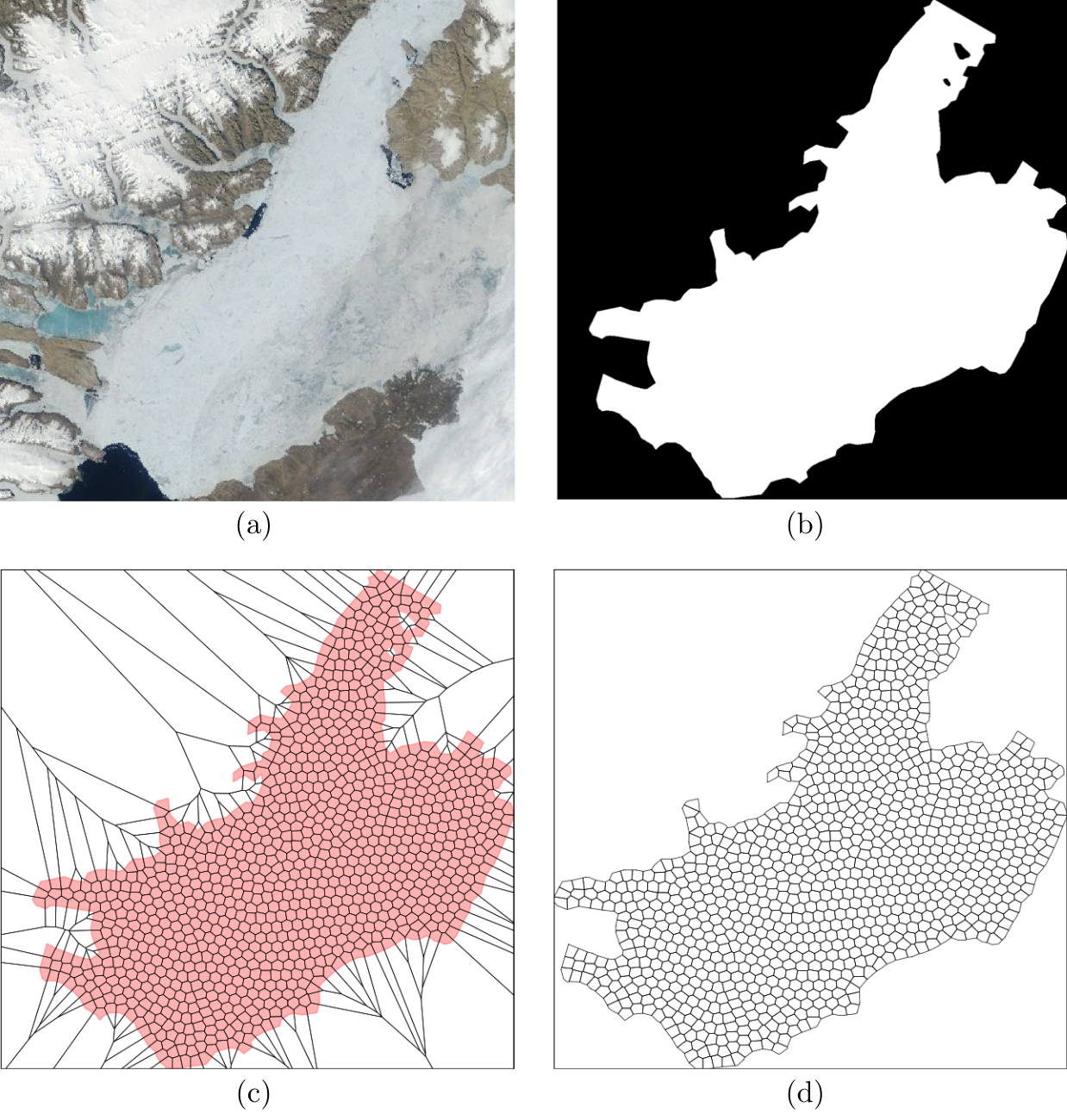}};
\end{tikzpicture}
\end{center}
\caption{Workflow for initializing polygonal ice floes from MODIS imagery. Image a is the MODIS imagery of the simulation domain, image b is a binary image reflecting ice extent used in the simulation, image c shows the entire set of polygons created by solving an optimal quantization problem with the ice extent outlined in red, and image d shows the final particle collection after clipping to the shape and extent of the input ice image. This set is intentionally a small number of particles (1000) for illustrative purposes.}
\label{fig:SDOT}
\end{figure}

To initialize our particle configurations, we leverage cloud-free MODIS imagery and concepts of optimal quantization from semi-discrete optimal transport \citep{xin2016centroidal,levy2018notions,bourne2018semi}. Using Otsu's Method \citep{Otsu1979} to threshold pixel intensities, we create a binary mask of sea ice in the image (see Figure \ref{fig:SDOT}b). We then treat this mask as a uniform probability distribution over the sea ice and find the best discrete approximation of this distribution using Lloyd's algorithm to solve the optimal quantization problem (see e.g., \citet{xin2016centroidal} and \citet{bourne2018semi}). As shown in Figure \ref{fig:SDOT}c, the result is a collection of points and polygonal cells over the entire domain. The polygonal cells form a power diagram, which is a generalization of a Voronoi diagram that enables cells to be weighted and thus have different sizes. Here, the cells are constructed so that they each have approximately the same overlap area with the sea ice (red region in Figure \ref{fig:SDOT}c). Within this framework, it is also possible to specify a distribution over cell-ice overlap area to generate particle configurations with specific floe size distributions (FSD).  While Voronoi diagrams are commonly used to construct polygonal DEM discretizations, we are unaware of other approaches that can randomly generate polygonal configurations with specified flow size distributions.

The final step in our initialization process is to identify the diagram cells that fill the ice extent (Figure \ref{fig:SDOT}c). Clipping the diagram cells by the ice extent can create concave, triangular, or small polygons shapes, which can affect the particle dynamics. Therefore, we define our ice particle geometries with the diagram cells that fall entirely within the ice extent, and take the cells that intersect the ice extent as our boundary particles. The final result is a set of polygons matching and filling the ice extent observed in the MODIS imagery (Figure \ref{fig:SDOT}d).

\changed{
\section{DEM Scaling Analysis}
\label{sec:scaling_approach}
Sea ice can accommodate relatively little deformation elastically.  Most large scale sea ice deformation therefore stems from fracture and motion along leads and larger linear kinematic features. As a result, large deformation rates tend to be concentrated in space and time.  Scaling analyses have been widely used to statistically quantify this behavior using both observations (e.g., \citet{Marsan2004, Rampal2008, Weiss2009b, Hutchings2011, Oikkonen2017}) and models (e.g., \citet{Girard2009,Girard2010,Dansereau2016,Rampal2019}).   In our results, we adapt the Delaunay triangulation approaches used by \citet{Oikkonen2017} and \citet{Rampal2019} to the DEM setting.  Scaling analyses are not commonly employed with DEM simulations. \changetwo{We have developed an approach that maps DEM particle positions to a Lagrangian mesh that can be used for computing strain rates with standard techniques from finite elements.  These strain rates can then be averaged over temporal and spatial windows of different sizes to characterize the intermittency and heterogeneity of the deformation. 

To be more specific, consider a strain rate field $\dot{\varepsilon}(x,t)$ that varies in space and time.  We can average the strain rate over some region $\mathcal{X}_\ell$ with size parameter $\ell$ and some time period $\mathcal{T}_\tau$ with length $\tau$, resulting in an average strain rate $\bar{\epsilon}_{\ell\tau}$.   The invariants of this average tensor can then be used to define a scalar total deformation rate $\dot{\varepsilon}_{\text{tot},\ell\tau}$ that also depends on the size of the averaging windows.   The dependence of $\dot{\varepsilon}_{\text{tot},\ell\tau}$ on the spatial window size $\ell$ and temporal window $\tau$ give insight into the localization of strain rate in space and time.  It can therefore be used to statistically compare the strain rate fields in a simulation to the intermittent and heterogeneous total deformation exhibited by real sea ice. \ref{sec:app:scaling}} provides a mathematically rigorous definition of the total deformation rate $\dot{\varepsilon}_{\text{tot},\ell\tau}$ as well as a description of how it can be efficiently computed from the output of a DEM simulation.    
} 

\section{Idealized Channel Simulation}
\label{sec:idealized_channel}
We use a simulation domain from \citet{Dansereau2017} as a baseline for \changed{testing our model's ability to simulate ice dynamics through a channel.} This geometry approximates the constriction from Kane Basin into Smith Sound within Nares Strait (see dimensions in Figure \ref{fig:damage-field}c). Following their simulation setup, we use a stagnant ocean field and a southward wind field starting at 0 m/s and increasing linearly to ${\sim}$22 m/s over 24 hours, which is then held constant. This wind approximates a storm passing \citep{Dansereau2017}. The model parameters for these different simulations are presented in Table \ref{table:model-params}. 

\begin{table}[!htbp]
    \caption{Model parameters used in simulations of sea ice advecting through the idealized channel and Nares Strait.}
    \begin{center}
    \begin{tabular}{|l|c|r|r|}
    \hline
    \bfseries Parameter & \bfseries Symbol & \bfseries Value & \bfseries Units \\ \hline
    Ice Density & $\rho_i$ & \SI{900.0} & \SI{}{kg/m^3} \\
    Air Density & $\rho_a$ & \SI{1.3} & \SI{}{kg/m^3} \\
    Ocean Density & $\rho_o$ & \SI{1027.0} & \SI{}{kg/m^3} \\
    Ice Young's Modulus & $E_i$ & $\SI{5.0e8}{}$ & $\SI{}{Pa}$ \\
    Ice Poisson's Ratio & $\nu_i$ & 0.3 &  \\ 
    Ice Thickness & $t_i$ & 1.0 & \SI{}{m} \\ 
    Wind Drag Coefficient & $C_a$ & \SI{1.5e-3} & \\ 
    Ocean Drag Coefficient & $C_o$ & \SI{5.5e-3} & \\ 
    Beam Radius Ratio & $r_{b}$ & 1.25e-2 & \\ 
    Beam Young's Modulus & $E_{b}$ & $\SI{5.0e8}{}$ & $\SI{}{Pa}$ \\ 
    Beam Poisson's Ratio & $\nu_{b}$ & 0.3 & \\ 
    Beam Damping Ratio & $\zeta_{b}$ & 0.7 & \\
    Mohr-Coulomb Internal Friction & $\mu$ & 0.7 & \\
    Mohr-Coulomb Tensile Strength & $\sigma_{N,t}$ & \SI{80.0e3} & \SI{}{Pa} \\
    Mohr-Coulomb Compressive Strength & $\sigma_{N,c}$ & \SI{-192.0e3} & \SI{}{Pa} \\
    Mohr-Coulomb Minimum Cohesion & $c_{min}$ & \SI{40e3} & \SI{}{Pa} \\
    Mohr-Coulomb Maximum Cohesion & $c_{max}$ & \SI{56e3} & \SI{}{Pa}\\
    Ridging Plastic Hardening & $k_{np}$ & \SI{928.0} & \SI{}{Pa} \\
    Ridging Plastic Drag & $k_r$ & \SI{26.1e3} & \SI{}{N/m} \\
    Ridging Friction Coefficient & $\mu_{r}$ & 0.3 & \\
    Ridging Damping Ratio & $\zeta_{r}$ & 1.0 & \\
    \hline
    \end{tabular}
    \label{table:model-params}
    \end{center}
\end{table}

\begin{figure}[h!]
\begin{center}
\begin{tikzpicture}
    \node[anchor=south west,inner sep=0] (image) at (0,8) {\includegraphics[width=1.0\textwidth]{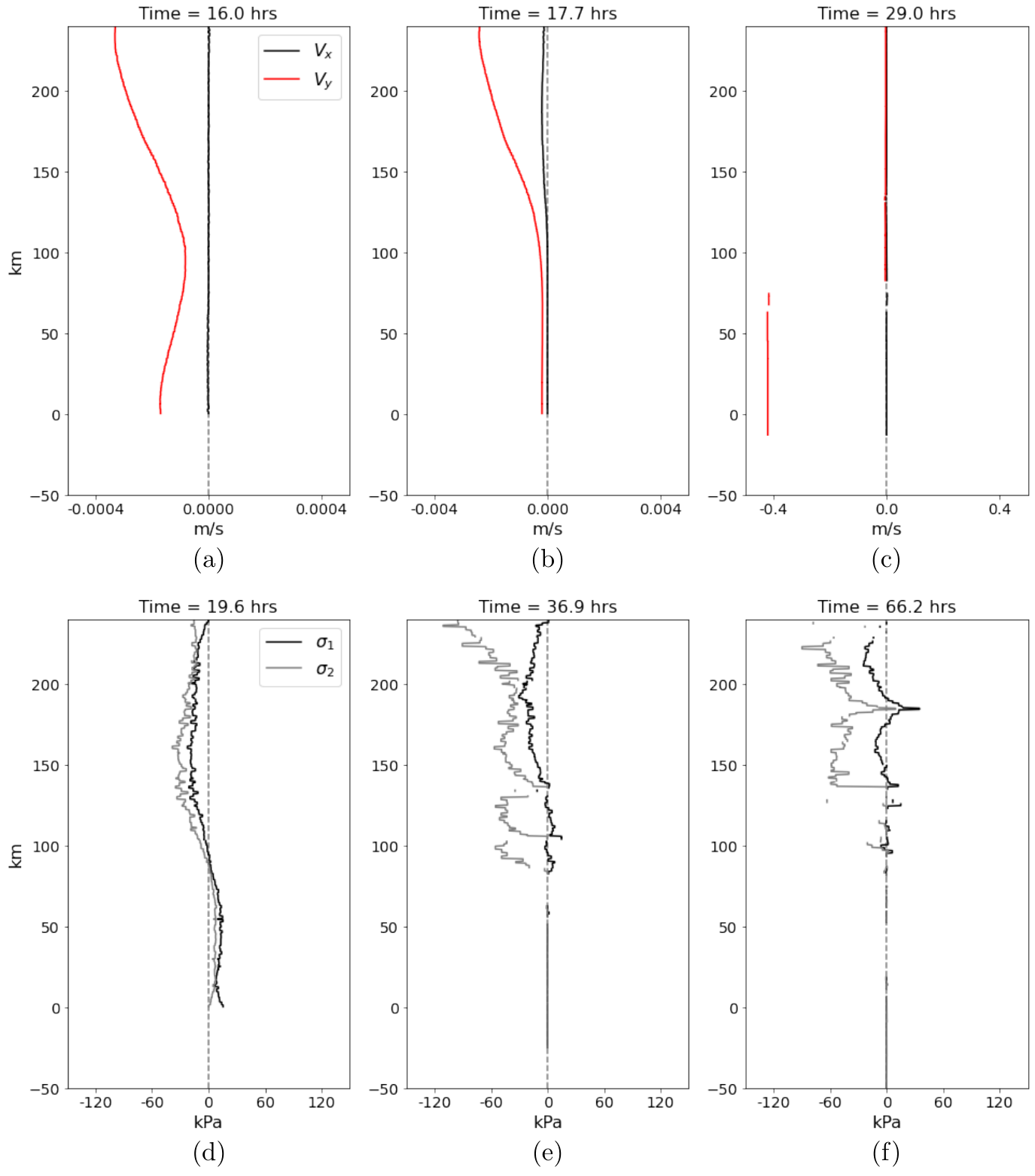}};
\end{tikzpicture}
\end{center}
\caption{Velocity and principal stress profiles measured along the central axis of the idealized geometry. The y-axis corresponds to the diagram in Figure \ref{fig:damage-field}c, where \changed{$\SI{0}{km}$} is the bottom of the channel geometry. Note that the velocity x-axis scale increases going from left to right.}
\label{fig:velocity-stress-profile}
\end{figure}
The domain starts as one contiguous piece of ice spanning the entire domain. The velocity profiles in Figure \ref{fig:velocity-stress-profile}a show how the ice initially has an hourglass-shape velocity profile along the central axis of the channel. This profile mimics the contours of the channel boundaries, and shows how the cohesive beams facilitate large scale deformations within the ice. The principal stress profiles in Figure \ref{fig:velocity-stress-profile}d also show a fairly continuous stress through the domain, with evidence of biaxial compression in the ice above the constricted region and biaxial tension below. The biaxial compression results from the ice being pushed into the convergent boundaries, whereas the biaxial tension results from the ice being pulled away from the divergent walls. 

\begin{figure}[h!]
\begin{center}
\begin{tikzpicture}
    \node[anchor=south west,inner sep=0] (image) at (0,0) {\includegraphics[width=0.9\textwidth]{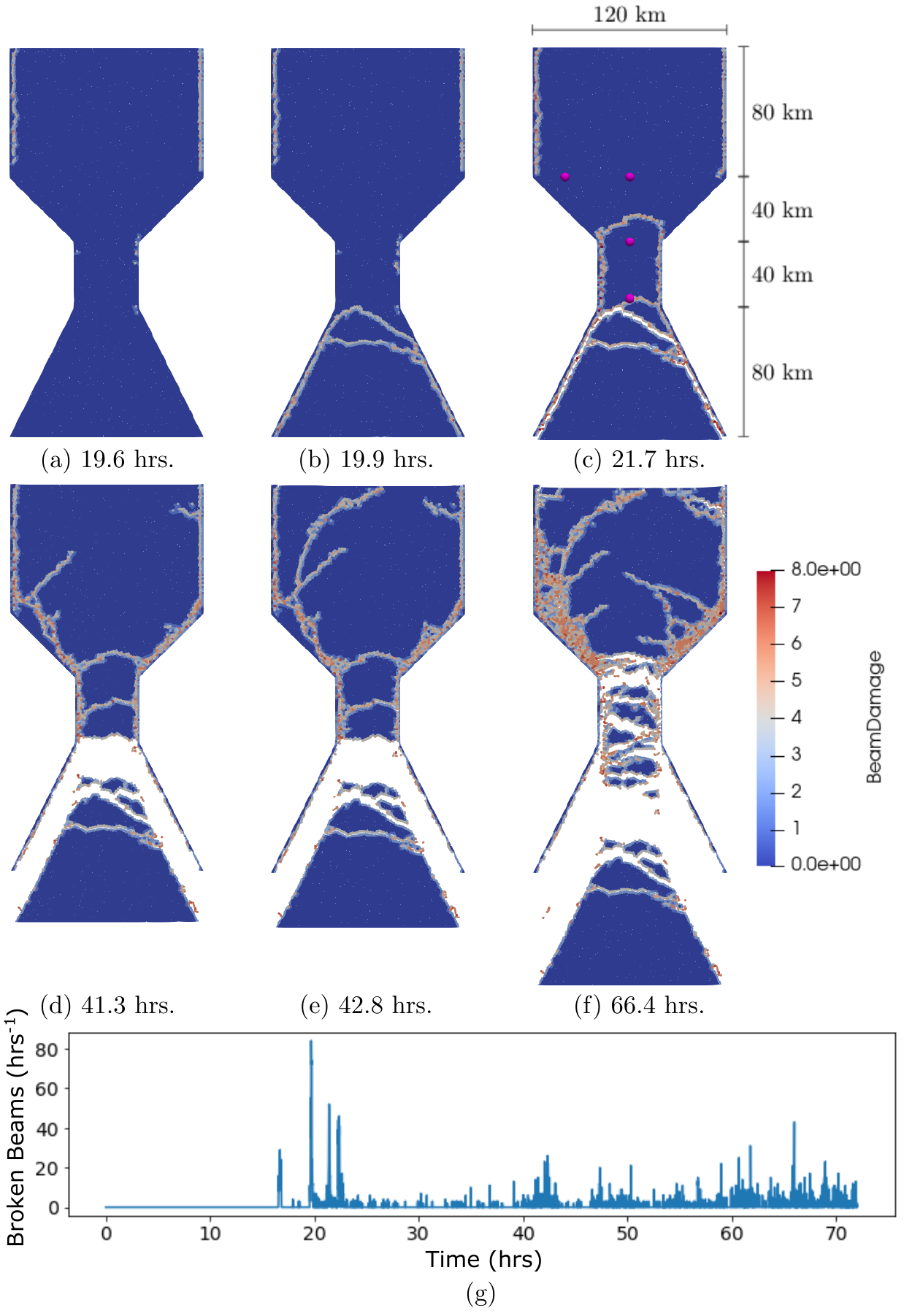}};
\end{tikzpicture}
\end{center}
\caption{\changed{Progression of ``beam damage''. Cracks initially form near corners along the boundaries, then propagate into the ice pack to form arches or linear features. Image g shows the intermittent rate of fracture throughout the simulation. The four points in image c correspond to the temporal scaling results in Section \ref{sec:idealized_channel}.}}
\label{fig:damage-field}
\end{figure}

\begin{figure}[h!]
\begin{center}
\begin{tikzpicture}
    \node[anchor=south west,inner sep=0] (image) at (0,0) {\includegraphics[width=\textwidth]{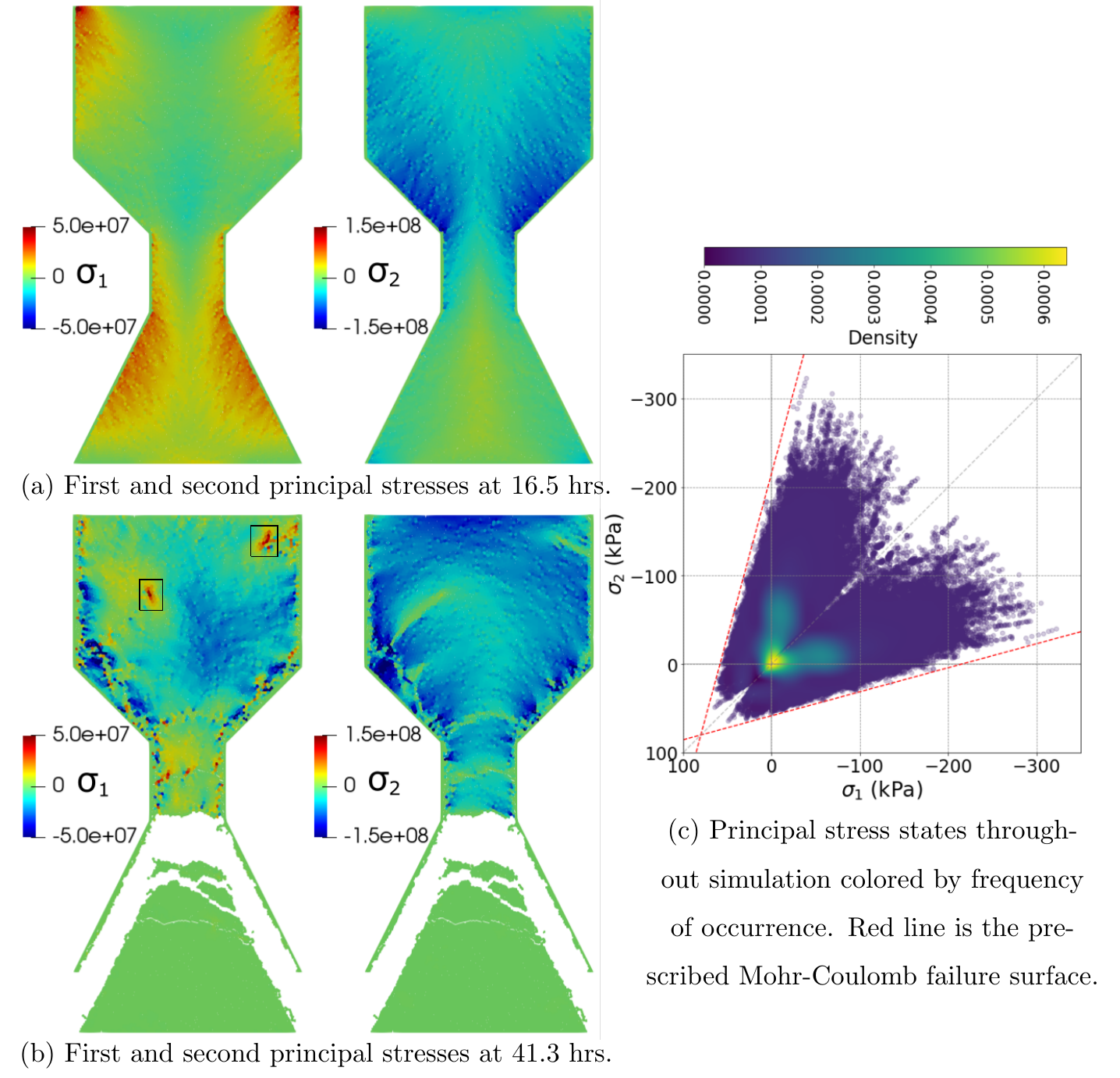}};
\end{tikzpicture}
\end{center}
\caption{Images a and b show the principal stress fields before and after fracture events. Note the different scales of $\sigma_2$ between a and b, as well as the two boxes in the $\sigma_1$ b image that show the location of crack tips moving through the ice. The damage field in Figure \ref{fig:damage-field}d corresponds to the same time as image b. Image c shows the stress states throughout the entire simulation, where the red dashed lines indicate a Mohr-Coulomb envelope with a cohesion stress of $c = \SI{56}{kPa}$, tension failure strength of $\sigma_{N,t} = \SI{-80}{kPa}$, and compression failure strength of $\sigma_{N,c} = \SI{192}{kPa}$. The coloring corresponds to the relative frequency of each stress value occurring throughout the simulation.}
\label{fig:stress-field}
\end{figure}

Cracks in the simulated ice are visualized with ``beam damage'', which is the number of bonds that have broken for each particle. Damage values of zero indicate particles with intact beams, whereas larger values indicate particles who have had several beams fail. The damage fields in Figure \ref{fig:damage-field}a-f and the damage time series in Figure \ref{fig:damage-field}g illustrate the highly intermittent ice fracture process. \changed{The beam damage rate in Figure \ref{fig:damage-field}g is analogous to the failure avalanches discussed in \cite{Girard2010} and is related to surface area of leads, and subsequently the fracture energy required to create those leads.} Many fractures originate along the boundaries and near corners (Figure \ref{fig:damage-field}a), as these features create stress concentrations in the ice. The first fractures occur at the top corners of the domain, where significant tension in $\sigma_1$ (Figure \ref{fig:stress-field}a) results from the wind drag pulling the ice downward. Eventually the beams in these regions fail, followed by linear cracks down the vertical walls. Once these cracks form the ice in the top region is no longer held in place by the boundaries and it starts to move. This is apparent in the increase in velocity in Figure \ref{fig:velocity-stress-profile}b for this region of the ice. Figure \ref{fig:damage-field}a shows that several fractures also originate near the corners of the thinnest channel section, which correspond to regions of large tensile or shear stresses in Figure \ref{fig:stress-field}. A closer inspection of Figures \ref{fig:damage-field}a and \ref{fig:stress-field}a shows that these individual fractures often connect with each other to form contiguous linear cracks along the boundaries. 

The next major event in the break up sequence is the formation of two cracks along the divergent angled boundaries, which eventually connect with each other near the exit of the channel and form an arch-shaped crack (Figures \ref{fig:damage-field}b and c). At this point the ice in the lower portion of the domain is completely separated from both the boundaries and the ice above the arch, and it begins to flow south in free-drift. This is clearly seen as the discontinuity in the velocity profile (Figure \ref{fig:velocity-stress-profile}c). This is an example of how the DEM is able to simulate the transition from one continuous piece of ice to multiple discrete pieces of ice. The reduced velocity in Figure \ref{fig:velocity-stress-profile}c above the arch show that the DEM approach is able to simulate how ice arches effectively plug the constricted region and do not allow the ice above them to move - an important aspect of ice arching in nature.

The $\sigma_1$ image in Figure \ref{fig:stress-field}b shows how the cracks propagating into the ice originate from fractures along the boundaries. These crack fronts are preceded by large tensile stresses (boxed regions in Figure \ref{fig:stress-field}b). These results are evidence that the model is able to capture cracks forming due to failure in tension, supporting observations of lead formation in sea ice \citep{Timco2010}. After this initial arch, the stresses above the constriction become more compressive as the ice is pushed against the convergent boundaries, whereas the stresses in the ice below the arch drop to zero because the ice is in free-drift. The ice within the channel experiences large shear stresses along the boundaries (Figure \ref{fig:stress-field}a) and ultimately fails (Figures \ref{fig:damage-field}b and c). These fractures then connect and form a clear arch in the convergent region above the channel (Figure \ref{fig:damage-field}c). This is followed by several linear features emanating from the vertical and convergent boundaries that sometimes connect to form a network of cracks surrounding regions of still-bonded particles--or floes.  Eventually the arch at the bottom of the channel fails and the ice within the channel breaks into smaller floes, which then move south. The top arch remains fairly stable, however the ice along the convergent boundaries continues to fail as it is crushed against the walls. 

Although not shown, several simulations were run and the trends described above match the general progression of all results. The arch in the simulation shown in Figures \ref{fig:velocity-stress-profile}-\ref{fig:stress-field} ultimately fails, however increasing the ice cohesion, $c$, above $\SI{64}{kPa}$ results in stable arches. Similar to \citet{Dansereau2017}, we do not attempt to identify appropriate cohesion values for these test cases as ice arch failure depends on a number of other factors including ice thickness and applied drag loads. Our goal is to illustrate that the bonded DEM model is a useful tool for estimating realistic sea ice dynamics \changed{within channel regions.}

\begin{figure}[h!]
\centering
\includegraphics[width=0.8\textwidth]{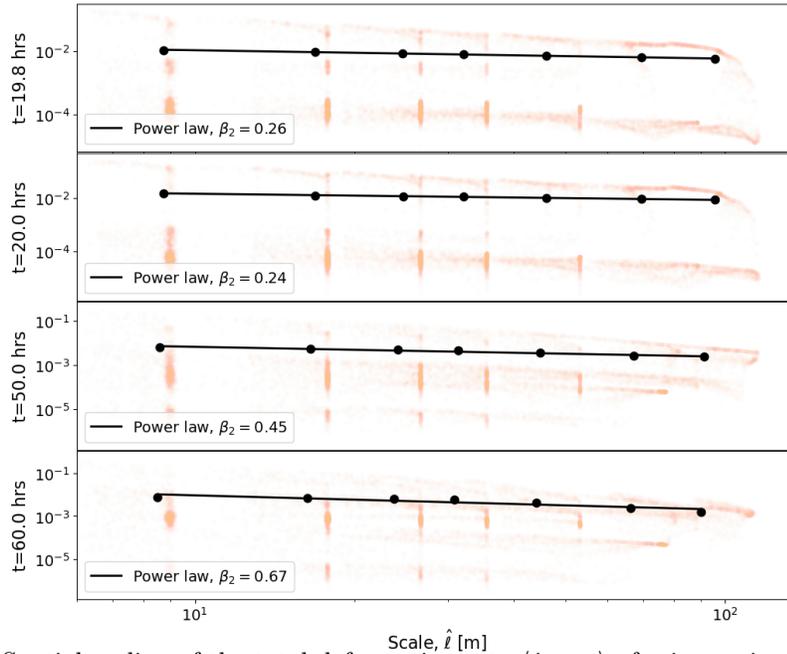}\vspace{-0.4cm}
\caption{\changed{Spatial scaling of the total deformation rate $\langle \dot{\varepsilon}_{\text{tot},\ell\tau}\rangle_x$ for increasing window sizes.}}
\label{fig:scaling:spatial}
\end{figure}

\begin{figure}[h!]
\centering
\includegraphics[width=0.8\textwidth]{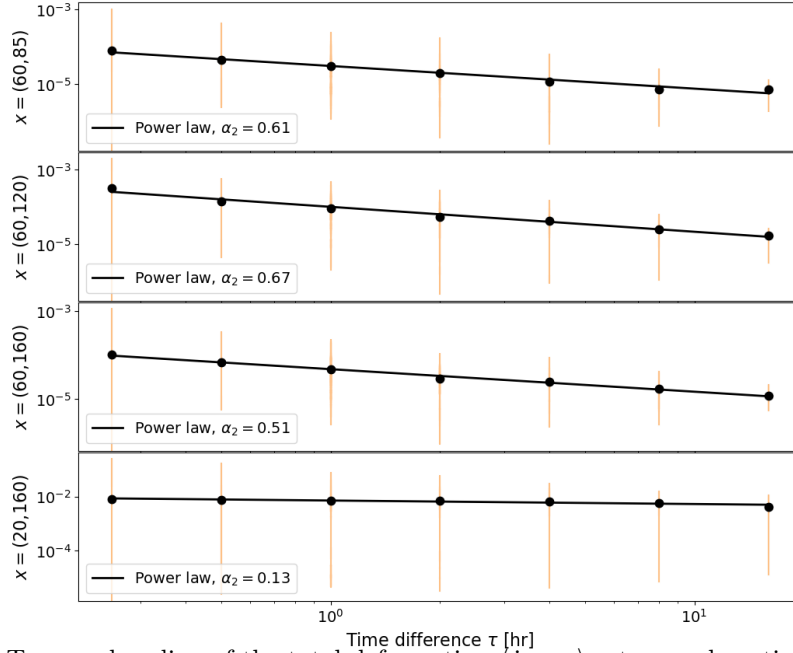}\vspace{-0.4cm}
\caption{\changed{Temporal scaling of the total deformation $\langle \dot{\varepsilon}_{\text{tot},\ell\tau}\rangle_t$ at several spatial locations within the channel domain. The locations of these points are highlighted in Figure \ref{fig:damage-field}c.}}
\label{fig:scaling:temporal}
\end{figure}

\changed{Two important characteristics of sea ice deformation are its heterogeneity (localization in space) and intermittency (localization in time) \citep{Weiss2007, Girard2009, Dansereau2016}, and recent studies have used these to assess how well numerical models capture the deformation of the modeled ice \citep{Dansereau2016, Girard2009}. Figure \ref{fig:damage-field}a-f shows how the DEM approach presented in this paper captures regions of highly-localized damage in the form of linear features that propagate through the ice, similar to what has been observed remotely \citep{Kwok2001} and in other modeling papers \citep{Dansereau2016, Girard2009}. Comparing these linear features with Figures \ref{fig:stress-field}a-c shows that these cracks coincide with regions of high tensile or compressive stresses, which make up a small portion of the overall ice stress states. Only 12.9\% of the stress states for all particles throughout the entire simulation fall outside of the high-frequency yellow and green regions in Figure \ref{fig:stress-field}c (probability density less than 0.0001).}

\changed{The time series in Figure \ref{fig:damage-field}g illustrates the sporadic evolution of ice damage throughout the simulation.} The drag loads in this simulation increase through the first 24 hours, and around 16.5 hours the ice begins to experience intermittent periods of large spikes in beam damage, followed by relatively calm periods of minimal break up. This cyclic behavior of stress building up in the ice followed by sudden relaxation through deformation is also seen in the work of \citet{Dansereau2016} and \citet{Weiss2017}. 

\begin{figure}[h!]
\begin{center}
\begin{tikzpicture}
  \node[anchor=south west,inner sep=0] (image) at (0,0) {\includegraphics[width=1.0\textwidth]{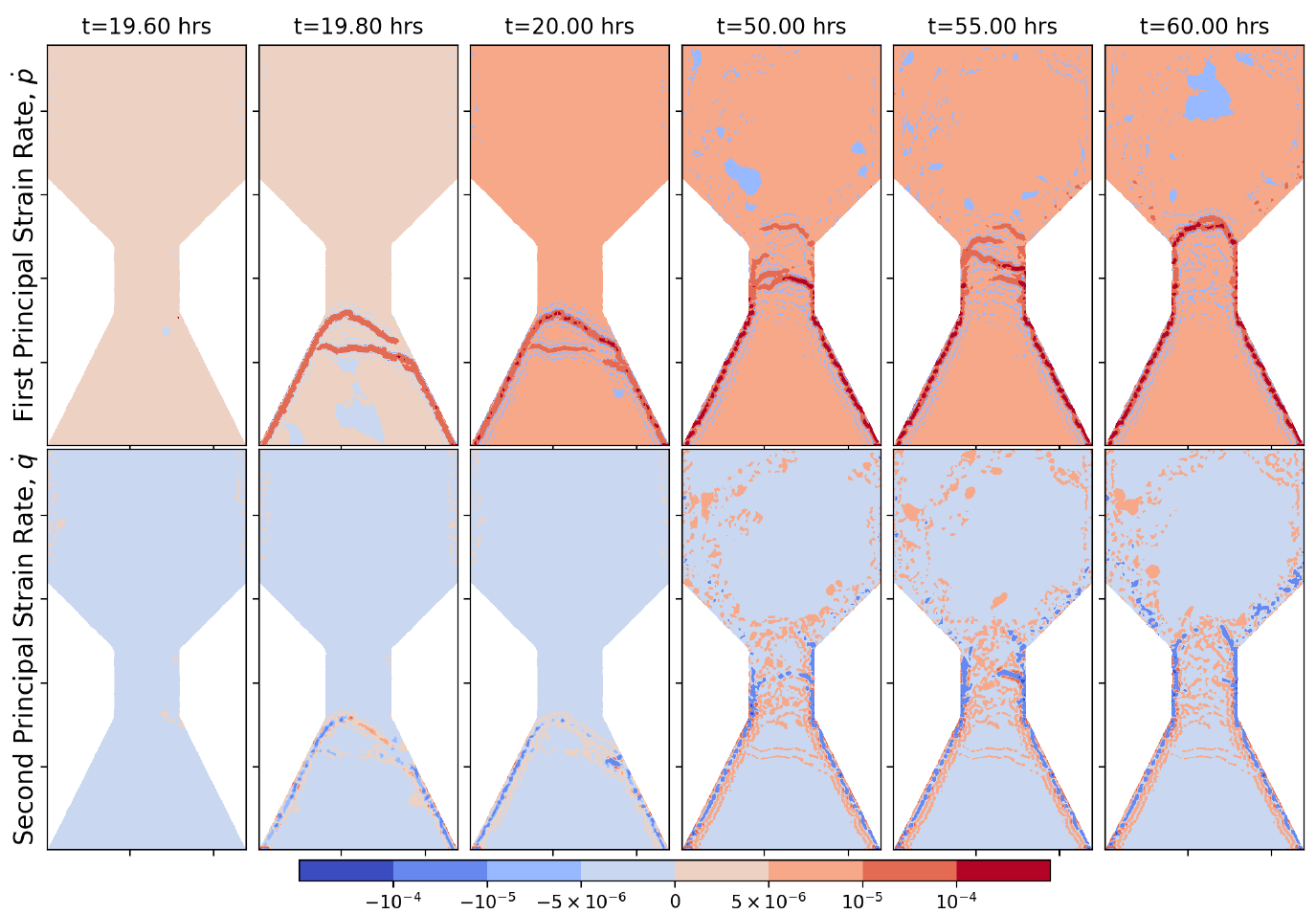}};
\end{tikzpicture}
\end{center}
\caption{\changed{Strain rates within the idealized channel simulation at different instances in time. Comparing these patterns with the beam damage fields in Figure \ref{fig:damage-field} indicates that the linear cracks coincide with regions of localized high strain rates. Note the arch shaped linear features that propagate up the channel throughout the break up process.}}
\label{fig:strain-localization}
\end{figure}

\changed{Figures \ref{fig:scaling:spatial} and \ref{fig:scaling:temporal} provide a spatio-temporal scaling analysis to further assess the heterogeneity and intermittency of dynamics in our simulation. The mean deformation rates (black dots) exhibit power law behavior (black lines), indicating the model captures localization of large strain rates in both space and in time. This is in agreement with scaling analyses of observed ice motion (see e.g., \citet{Marsan2004,Oikkonen2017}) as well as other modeling results (see e.g., \citet{Girard2009,Dansereau2016,Rampal2019}). The values of $\beta_2$, which are larger at later times, are in agreement with the damage fields in Figure \ref{fig:damage-field} and the strain rates in Figure \ref{fig:strain-localization}.  Initially the ice has relatively homogeneous strain rates, except for a few localized arching events, but the strain rates are more heterogeneous at later times when the ice has broken into many small floes. The temporal scaling coefficients are larger for points in and below the neck of the channel, which indicates strong temporal localization of strain rates in these areas. This makes intuitive sense: these regions experience a short period of high strain rates during the initial fracture event and then are in relatively free drift.}

\changed{
Figure \ref{fig:strain-localization} complements the quantitative scaling analysis with a visual representation of the principal strain rates.  The velocities of the DEM particles are mapped onto a Delaunay triangulation of the particle centroids at $t=0$, which allows the strain rate tensor to be computed over the cells in the triangulation.   The strain rates are localized in the same regions that experience large damage rates (see Figure \ref{fig:damage-field}).  Bands of compressive strain rates (negative values) can also be seen on either side of large tensile strain rates (positive values), indicating that the arches are supporting the ice above.
}

We feel the results from the idealized channel simulations show how the bonded DEM approach is able to capture the salient features of ice advecting through a constriction and the subsequent jamming, \changed{as well as important deformation characteristics (heterogeneity and intermittency) seen in real sea ice.} Next, we apply this same model to the Nares Strait geometry and estimate a distribution of floe areas and the amount of ice flowing out of Kane Basin into Smith Sound.

\section{Nares Strait Simulation}
\label{sec:nares-strait}
In our Nares Strait simulations we once again adopt the linearly-increasing wind current and stagnant ocean current used in \citet{Dansereau2017}. The wind field is oriented down channel starting at 0 m/s and increasing to ${\sim}$22 m/s over 24 hours, which is then held constant through 72 hours. As noted by \citet{Dansereau2017}, ice motion through Nares Strait is believed to be primarily driven by winds flowing south between Ellesmere Island and Greenland. The model parameters used in these simulations are similar to those in Table \ref{table:model-params}, except for the number of particles. Our model domain is a reduced region of Nares Strait focused on Kane Basin, and we use MODIS imagery from June 28, 2003 to initialize the ice extent (see section \ref{sec:particle-init} and Figure \ref{fig:SDOT}a). We chose the June 28, 2003 ice state because the clarity of the MODIS imagery before and after the arch fails provides a useful comparison. The resultant particle set has $8682$ polygonal ice particles, and $695$ stationary boundary particles. Although not shown here, we created additional particle set with \changed{more and less} ice particles and found very similar results, suggesting that the $8682$ particle set is able to capture the salient dynamics. 

\begin{figure}[h!]
\begin{center}
\begin{tikzpicture}
    \node[anchor=south west,inner sep=0] (image) at (0,0) {\includegraphics[width=0.95\textwidth]{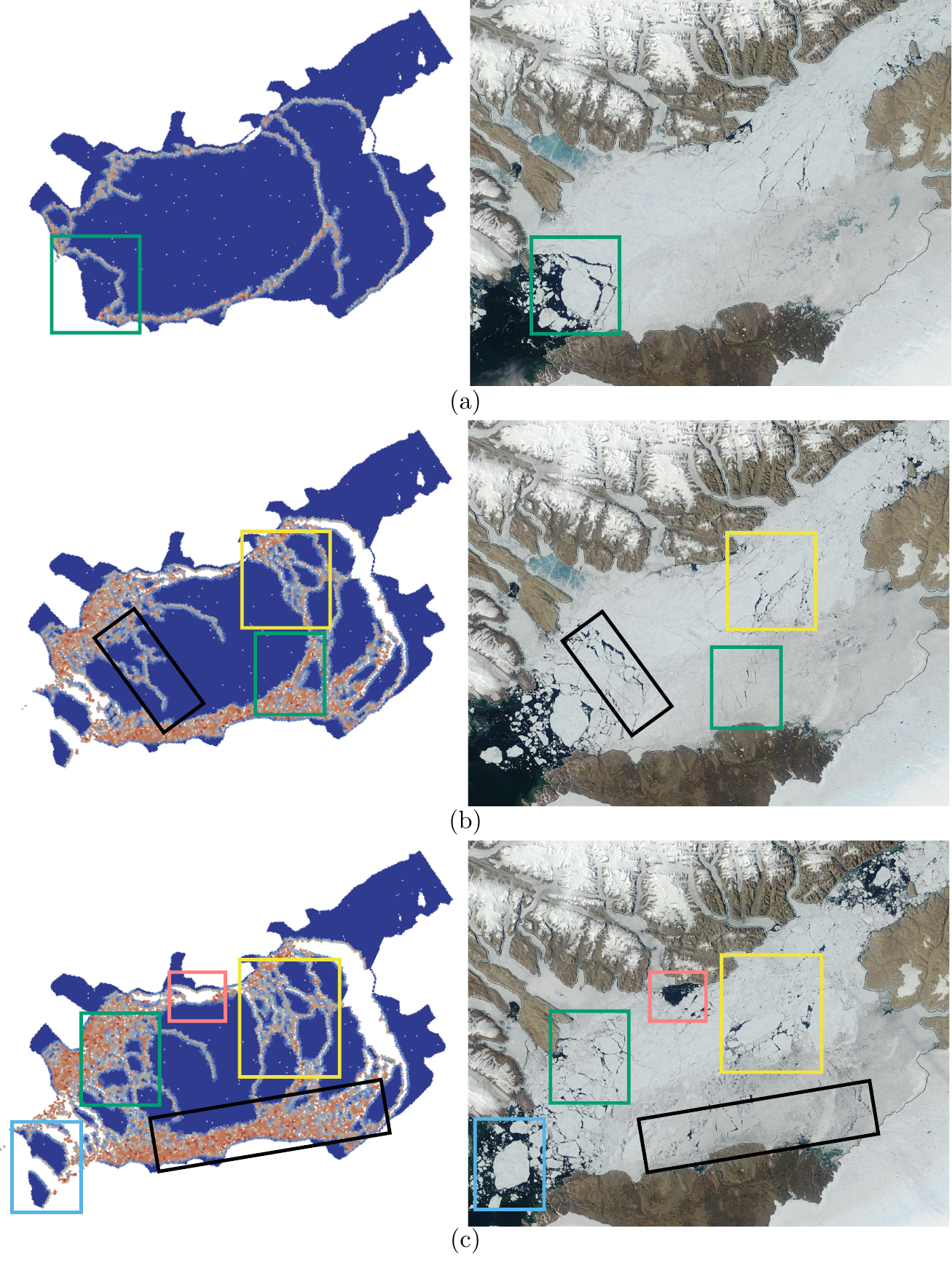}};
\end{tikzpicture}
\end{center}
\caption{Comparison of ``beam damage'' throughout the Nares Strait simulation with MODIS images of the actual ice break up. The colored boxes indicate regions of interest where the model captures features of the \changed{actual} ice break up. The colorbar for the simulated results are the same as in Figure \ref{fig:damage-field}. The MODIS images are courtesy of NASA Earth Observing System Data and Information System (EOSDIS).}
\label{fig:NS-comparison}
\end{figure}

Our model uses synthetic wind and ocean loads, as well as a uniform ice thickness of 1 m, meaning the driving forces and ice conditions in the model do not precisely match the conditions in the real Nares Strait. Due to these discrepancies, we do not expect an exact match between model and observations, and therefore provide a qualitative comparison in Figure \ref{fig:NS-comparison} as an illustration of how the bonded DEM model is a useful tool for simulating and studying ice dynamics \changed{within channel domains.} Despite the aforementioned differences, there are similarities between the model and observations. Figure \ref{fig:NS-comparison}a shows a rounded fracture upstream of the initial arch, resulting from tensile failure near the right edge of the arch that propagates into the ice. This arch-like fracture is clearly seen as one of the first major break up events in the corresponding MODIS image. As the break up progresses to Figure \ref{fig:NS-comparison}b, additional fractures form upstream of these initial arch-like cracks, which is captured by the model (black boxes). The ice in the yellow boxes has begun to break up further, and a series linear of cracks have started emanating from the coastline as the ice is crushed and sheared against the land (green boxes). 

At this point in the simulation there are multiple cracks bisecting the channel and long fractures along the boundaries that effectively separate the ice in the side inlets and channels from the ice in Kane Basin. After a period of time the cracks along the boundaries accumulate more damage as the ice is crushed against the coastline. Eventually the ice in the middle of the channel is no longer bonded to the boundaries and it begins to flow into Smith Sound. Similarly, we see that the observed ice also begins to move towards Smith Sound, but not uniformly. The ice moves fastest within a linear region extending from the exit of Kennedy Channel to the entrance to Smith Sound. The ice to the east of this region moves slower--particularly the ice near Humboldt Glacier. The model contains multiple cracks that separate this portion of the ice from the main channel, which is predominantly landfast. \changetwo{Landfast ice is also modeled in other regions, especially} in the fjords, inlets, and channels off of Nares Strait, which is also observed in the simulations of \citet{Dansereau2017}, the RADARSAT observations of \citet{Yackel2001}, and the estimated strains in \citet{Parno2019}. 

\changetwo{The ice continues to break up as it advects out of Kane Basin (Figure \ref{fig:NS-comparison}c), and considerable break up occurs along the southern coastlines that form the constriction. The model is able to capture the ice crushing (black boxes) and breaking up into floe-like objects (green boxes) in regions similar to the MODIS imagery. Interestingly, the model also captures the formation of an open-water region (pink boxes) as the ice is sheared away from the western coastline. The ice near the exit of Kennedy channel continues to break up into many large floes (yellow boxes). Eventually the southern arch fails completely\changed{, and our model produces several floe-like objects exiting Kane Basin}, which is also clearly seen in the corresponding MODIS image (light-blue boxes).

One major difference between the model and observations is that the simulation produces a stable arch where Kennedy Channel enters Kane Basin. This arch restricts ice from advecting into and ``refilling'' Kane basin, which results in the large open water region near the top of the basin. This is not observed in the MODIS imagery and this model-reality mismatch is likely a result of the model initial conditions and wind direction. The model starts with 100\% ice concentration in Kennedy channel with ice that is also bonded to the sides of the channel. This landfast ice likely overestimates the strength of the ice in the region, creating conditions where a stable arch can form. The MODIS image in Figure \ref{fig:NS-comparison}a indicates that the ice in Kennedy Channel has clear areas of open water, and there does not appear to be significant regions of landfast ice, thus allowing more of the ice to advect into Nares Strait. In \citet{Parno2019}, the ice in Nares Strait was also observed to flow in from Kennedy channel towards Humboldt Glacier. Despite there being no stable arch in the MODIS imagery, this modeled arch closely matches an arch in the Nares Strait simulation of \citet{Dansereau2017} using similar conditions (see Figure 6c 72 hour column in \citet{Dansereau2017}).}

\begin{figure}[h!]
\begin{center}
\begin{tikzpicture}
    \node[anchor=south west,inner sep=0] (image) at (0,0) {\includegraphics[width=\textwidth]{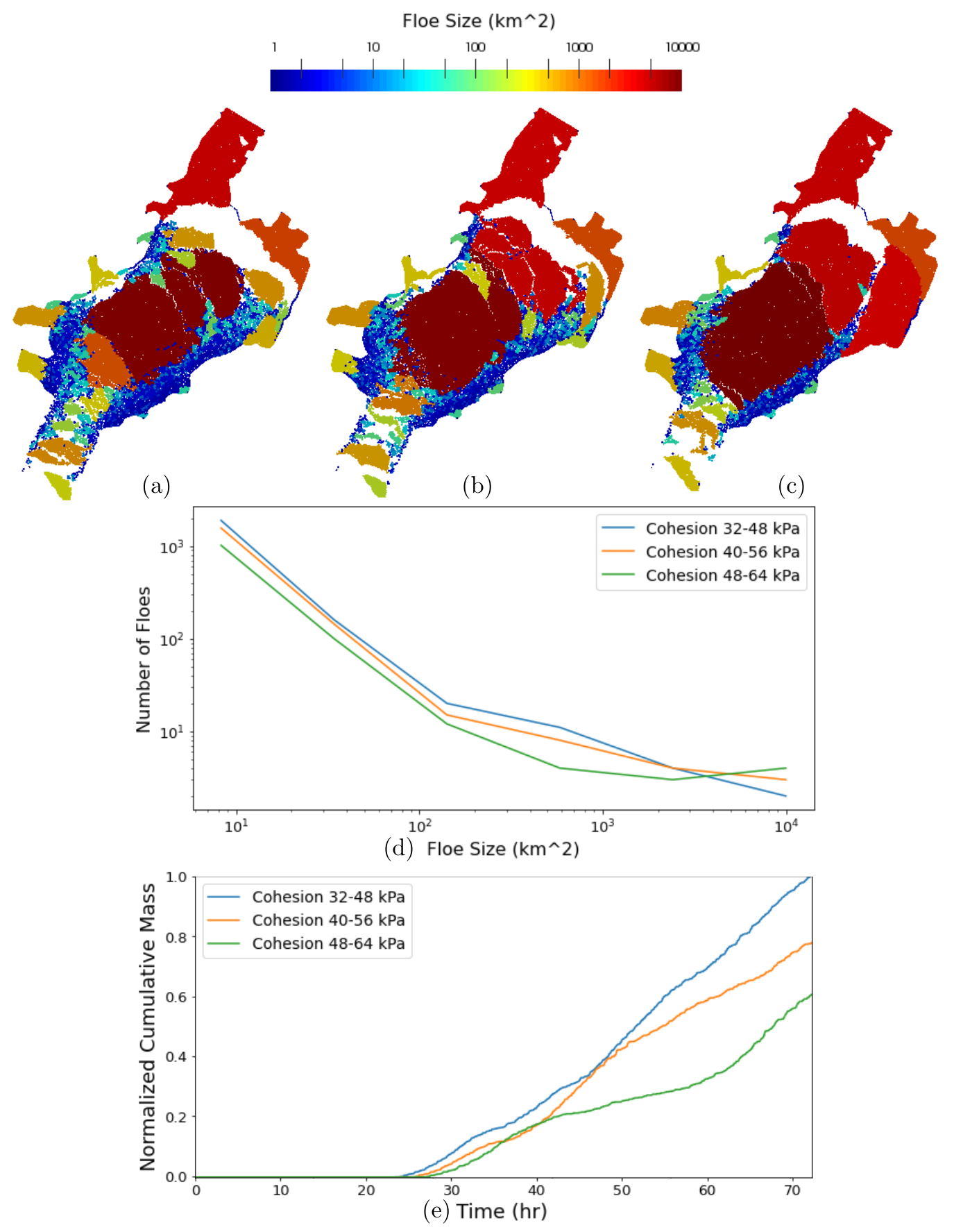}};
\end{tikzpicture}
\end{center}
\caption{Floe size area (km$^2$) for three different simulations after 72 hours - (b) $c_{min} = \SI{32}{kPa}$ and $c_{max} = \SI{48}{kPa}$, (c) $c_{min} = \SI{40}{kPa}$ and $c_{max} = \SI{56}{kPa}$, (d) $c_{min} = \SI{48}{kPa}$ and $c_{max} = \SI{64}{kPa}$. The results in b correspond to the same simulation in Figure \ref{fig:NS-comparison}. Image c is the comparison of cumulative ice mass export ice leaving Kane Basin into Smith Sound (approximately the location of the initial arch in Figure \ref{fig:NS-comparison}a).}
\label{fig:NS-floe-area-export}
\end{figure}
We quantify individual floes as regions of particles that are still connected to each other through cohesive beams. Varying the material cohesion parameter affects the amount of break up in the ice, which therefore affects the size distribution of the simulated floes leaving the channel. Figure \ref{fig:NS-floe-area-export}d compares distributions of floe area from three different simulations with different cohesion ranges after 72 hours. Similar to \citet{Dansereau2017}, lower cohesion results in more break up, as indicated by the larger number of small floes for lower cohesion distributions in Figure \ref{fig:NS-floe-area-export}d. Although we are unaware of any observed floe size distributions for Nares Strait in the literature, the area distributions follow the general trend of few large floes and many small floes, which match general observations from the field \citep{Weiss2004}. A significant percentage of these small floes are particles whose bonds have entirely failed through crushing against the coastlines, which can be seen as the large blue regions in Figure \ref{fig:NS-floe-area-export}a, b, and c. The size of these highly-damaged regions appear to increase in size as cohesion values decrease, which reflects weaker ice crushing more readily against boundaries than stronger ice.

Variation in how much the ice breaks apart directly affects the mass export out of Nares Strait. Figure \ref{fig:NS-floe-area-export}e shows the normalized ice mass exiting Kane Basin into Smith Sound for the three simulations above. The results are normalized by the largest mass export at $T = 72$ hours for the $c_{min} = \SI{32}{kPa}$ and $c_{max} = \SI{48}{kPa}$ case in order to show general trends in the simulated ice mass export for the region. We assume a uniform ice thickness, and therefore it is misleading to directly compare to the simulated ice mass to observations of ice with varying thickness. The ice in all three simulations start to leave Kane Basin at roughly the same time and same rate, however the final mass exports are significantly different, with lower cohesion values corresponding to larger mass export. The lower cohesion ice breaks into many small floes, which are able to flow out of the basin at a higher rate than the stronger ice, which remains consolidated in larger floes. These results indicate that weaker ice can lead to earlier outflow and more overall ice moving through Nares Strait, which supports the findings of \citet{Dansereau2017} and \citet{Moore2021}. These results also suggest the bonded DEM could be a useful approach for studying the increase in ice export seen in recent years through Nares Strait \citep{Moore2021}, particularly as increasingly realistic ice thickness, wind forcing, and other variables are incorporated into future versions of the model.

\section{Discussion and Conclusions}
\label{sec:conclusion}

\changed{We present a bonded DEM model that uses the cohesive beam model and a non-local Mohr-Coulomb failure approach to simulate sea ice dynamics. We use an idealized channel domain and a Nares Strait domain to illustrate how the model can deform continuous ice and subsequently fracture it into many disparate floes. Figures \ref{fig:velocity-stress-profile}a, \ref{fig:velocity-stress-profile}d, and \ref{fig:stress-field}a show how the model can simulate continuous velocities and stresses throughout the ice that account for boundary effects and stress concentrations. Figure \ref{fig:stress-field}b shows that once failure occurs, large tensile stresses often precede the crack tips as they propagate through the ice, which matches observations of lead formation in nature \citep{Timco2010}. The results in Figures \ref{fig:velocity-stress-profile}c, \ref{fig:damage-field}, and \ref{fig:NS-comparison} show how the model produces many of the salient features of ice advecting through constricted regions-namely jamming, arch-shaped fractures, and ice crushing against solid boundaries. The scaling analyses presented in Figures \ref{fig:scaling:spatial} and \ref{fig:scaling:temporal} illustrate how our bonded DEM simulations exhibit heterogeneity and intermittency in the resultant ice deformation. These metrics have been used to validate continuum sea ice models in the past, but to the best of our knowledge, have not previously been applied to DEM models of sea ice.}

\changed{Section \ref{sec:failure_model} and the work of \citet{Andre2013} highlight that local per-beam failure models used in previous DEM studies can fail to capture continuous fracture paths in elastic brittle materials. These methods do not consider the fracturing events occurring near each other within the ice, and therefore can exhibit fragmenting behavior. We addressed this issue with a non-local failure model that considers the stress and fractures occurring within a small region around each particle. If the particle's stress state violates a Mohr-Coulomb criteria then the model selectively chooses which bonds to break at that instance in time, and therefore avoids the fragmenting behavior observed by \citet{Andre2013}. In addition, our bond clipping method encourages tensile crack growth, matching observations of ice.}

\changed{Comparing the Nares Strait simulation with the MODIS images in Figure \ref{fig:NS-comparison} shows the \changed{potential for using this model to simulate} real world scenarios. The model is able to qualitatively capture many of the salient features, including how the southern arch fractures into multiple large floes, and the development of multiple arch-like fractures upstream within Kane Basin. The model also accurately simulates landfast ice in the channels and fjords off of the Basin and near Humboldt Glacier, similar to the observations of \citet{Yackel2001}. Figure \ref{fig:NS-floe-area-export} shows how the modeled ice fractures into different sized floes near the exit of Kane Basin into Smith Sound, similar to the observed ice in Figure \ref{fig:NS-comparison}a. As expected, we see a correlation between weaker ice, earlier failure of the ice arches, and increased ice export out of the strait.}

\changed{The idealized channel simulations allow us to compare our DEM results with the different continuum approaches used to simulate ice advecting through similar geometries. Both \citet{Dumont2009} and \citet{Rasmussen2010} used models based on the EVP rheology, and \citet{Dumont2009} showed that it is possible to capture stable ice bridges in a channel by modifying the eccentricity of the EVP elliptical yield curve. However, \citet{Rasmussen2010} noted that due to the isotropic assumption in the EVP model, it may be unsuitable for simulating ice in Nares Strait because the complex coastline affects the ice stress state at much smaller scales than $100$ km. Alternatively, \citet{Dansereau2017} used the Maxwell elasto-brittle (Maxwell-EB) model, which tracks strain induced damage in the ice to approximate the location of leads and cracks.}

\changed{Our} results in Figures \ref{fig:velocity-stress-profile}, \ref{fig:damage-field}, and \ref{fig:stress-field} match the simulated results in \citet{Dansereau2017} remarkably well considering the differences in modeling approaches. We believe this is one of the strengths in our approach. While DEM models are known to be well-suited for MIZ simulations \citep{Damsgaard2018}, where continuum sea ice methods may suffer in accuracy, we believe the results in Sections \ref{sec:idealized_channel} and \ref{sec:nares-strait} also indicate that the DEM can \changed{qualitatively match the continuum-like behavior captured with the Maxwell-EB model, as well as subsequent complex fracture events, \changetwo{for sea ice flowing through channels}. In addition, the spatial and temporal analyses indicate that the bonded DEM is able to capture important deformation properties of sea ice\changetwo{, like spatial heterogeneity and temporal intermittency.} This suggests that DEM models have the potential to capture sea ice} behavior across contiguous, fractured, and completely broken regimes. \changed{We do not attempt to definitively state when and where DEM models should be used instead of continuum models, as both approaches have utility in the sea ice modeling landscape. Instead, we aim to show that the bonded DEM approach can capture continuum-like behavior within consolidated ice, as well as the transition to highly-discontinuous ice after failure. Future work will continue to validate the model results against observations of real ice, in non-channel domains, and across a range of spatial and temporal scales.}

Despite the qualitative agreement between our model results, the \citet{Dansereau2017} results, and satellite observations, there are several areas where the DEM model could be improved. First and foremost, assimilating more observational data into the model could improve accuracy. For example, we used wind speeds that approximate a large idealized storm passing through the idealized channel and Nares Strait. Actual winds were \changed{s}lower and more complex. As a result we see much larger displacements in that simulation than after 72 hours in the MODIS imagery. This uniform wind load and the stagnant ocean load vastly oversimplify the drag loads acting on the real ice. Incorporating more accurate wind and ocean data could improve the accuracy of the model. In addition, infusing additional data products such as SAR imagery can inform future simulations with a better understanding of the ice type (first-year or multi-year), thickness, or existing flaws, which can significantly change the ice properties. Future simulations will assimilate more data, as it's available.

At this point our model does not evolve any thermodynamics or change the ice thickness throughout the simulation. \citet{Hibler2006} states that the Nares Strait arch may become stronger due to thermodynamic processes, which our model ignores, and could be a source of mismatch between the simulated results and observations. However, the time scales of these DEM simulations are quite short - on the order of several hours or a few days. Effects such as thermodynamic thickening likely play a smaller role in the dynamics over these short timescales. However, mechanical thickening could play an important role in these regional scale simulations, particularly in the large crushing regions in Figures \ref{fig:NS-comparison} and \ref{fig:NS-floe-area-export} where the ice in Nares Strait would likely become thicker due to ridging. In fact these same regions become thicker in the Nares Strait simulations in both \citet{Dumont2009} (Figure 13) and \citet{Dansereau2017} (Figure 11a). Future DEM studies will vary \changed{ice particle thicknesses to investigate how thickness} affects arch stability, and how it relates to earlier arch break up and greater export out of the strait. 

A known limitation with bonded DEM or lattice spring methods is the need to calibrate local model parameters \citep{Nguyen2019}. Often times setting the bond's properties such as Young's Modulus, or failure strengths to the macroscopic values of a particular material do not yield realistic results. The extra step of calibrating these parameters to achieve realistic elastic and fracture behavior can be time consuming, and does not guarantee accurate macroscopic behavior. Future work may incorporate an optimization routine to learn the appropriate model parameters from the mismatch between model output and satellite observations. Alternatively, the use of non-local distinct lattice spring \citep{Andre2019}, or peridynamic models \citep{Davis2021, Silling2005} could avoid the need for time intensive calibration studies, and facilitate using real-world values for the model parameters. 

\changed{As sea ice models continue to develop towards forecasting dynamics on tactically-relevant scales, the ability to model explicit leads and cracks in the ice may prove critical to the overall utility of the ice forecasts. Future studies will look at how well the bonded DEM method presented here can capture dynamics across a range of spatial scales, including those relevant to navigation and shipping. We feel that the bonded-DEM with a non-local failure model shows promise as a useful tool to provide estimates of compression, deformation, and lead formation, thereby filling the gaps in current operational ice products identified by \citet{Mariner2019}.}

\section{Open Research}
Information on the ParticLS software library is included in \citet{Davis2021}, and the parameters necessary to reproduce these ParticLS simulations are described in the text and in Table 1. MODIS imagery were provided by the NASA Worldview application (https://worldview.earthdata.nasa.gov/).

\acknowledgments
We would like to thank the reviewers for their careful review and constructive feedback, which greatly improved our article. This research was funded as part of the U.S. Army Program Element 060311A, Ground Advanced Technology task for Sensing and Prediction of Arctic Maritime Coastal Conditions, through the Office of Naval Research SIDEx program, grant N000142MP001, and an Office of Naval Research MURI grant N00014-20-1-2595. We acknowledge the use of MODIS imagery provided by the NASA Worldview application (https://worldview.earthdata.nasa.gov/), part of the NASA Earth Observing System Data and Information System (EOSDIS). 

\bibliography{references} 

\appendix 

\changetwo{
\section{Details of Scaling Analysis.}\label{sec:app:scaling}
\subsection{Mathematical Formulation.} Consider a strain rate tensor $\dot{\varepsilon}(x,t)$ that varies with spatial location $x$ and time $t$.  This tensor could be derived from observations of sea ice velocities or from the output of a sea ice model.  Scaling analyses consider spatio-temporal averages of this pointwise strain tensor, where the average is taken over spatial subdomains $\mathcal{X}_\ell(x^\ast)\subset\mathbb{R}^2$ defined by a length scale $\ell$ and time intervals $\mathcal{T}_\tau(t)\subset \mathbb{R}^1$ defined by a timescale $\tau$.  Mathematically, the average strain rate tensor is given by
\begin{linenomath*}
\begin{equation}
    \bar{\epsilon}_{\ell\tau}(x^\ast,t^\ast) = \frac{1}{|\mathcal{X}_\ell(x^\ast)| |\mathcal{T}_\tau(t^\ast)|}\int_{\mathcal{X}_\ell(x^\ast)} \int_{\mathcal{T}_\tau(t^\ast)} \dot{\varepsilon}(x,t)\, dt\, dx,
    \label{eq:avg_sr}
\end{equation}
\end{linenomath*}
where $|\mathcal{X}_\ell(x^\ast)|$ and $|\mathcal{T}_\tau(t^\ast)|$ denote the area of $\mathcal{X}_\ell$ and length of $\mathcal{T}_\tau$, respectively.   From this average strain rate tensor, the total deformation rate $\dot{\varepsilon}_{\text{tot},\ell\tau}$ can be computed as 
\begin{linenomath*}
\begin{equation}
    \dot{\varepsilon}_{\text{tot},\ell\tau} = \sqrt{\dot{\varepsilon}_{\text{d},\ell\tau}^2 + \dot{\varepsilon}_{\text{s},\ell\tau}^2},
    \label{eq:total_def}
\end{equation}
\end{linenomath*}
where $\dot{\varepsilon}_{\text{d},\ell\tau}$ and $\dot{\varepsilon}_{\text{s},\ell\tau}$ are the divergent and shear components of the average strain rate, defined as 
\begin{linenomath*}
\begin{equation}
    \begin{aligned}
    \dot{\varepsilon}_{\text{d},\ell\tau} & = \bar{\varepsilon}_{\ell\tau,xx} + \bar{\varepsilon}_{\ell\tau,yy}\\
    \dot{\varepsilon}_{\text{s},\ell\tau} &= \sqrt{(\bar{\varepsilon}_{\ell\tau,xx}-\bar{\varepsilon}_{\ell\tau,yy})^2 + (\bar{\varepsilon}_{\ell\tau,xy} + \bar{\varepsilon}_{\ell\tau,yx})^2}.
    \end{aligned}
\end{equation}
\end{linenomath*}
Notice that the total deformation rate $\dot{\varepsilon}_{\text{tot},\ell\tau}$ is a function of position $x$ and time $t$ but also has a dependence on the length scale $\ell$ and timescale $\tau$.

The relationship of $\dot{\varepsilon}_{\text{tot},\ell\tau}$ with scales $\ell$ and $\tau$ provides insight into the structure of the deformation field.  Many studies have observed that, when averaged over all positions $x$, the total deformation rate has a power law relationship with $\ell$ (e.g., \citet{Marsan2004, Hutchings2011}), so that 
\begin{linenomath*}
\begin{equation}
\langle \dot{\varepsilon}_{\text{tot},\ell\tau}\rangle_x \approx \beta_1(\tau) \ell^{-\beta_2(\tau)},\label{eq:spatial_scaling}
\end{equation}
\end{linenomath*}
where $\langle \cdot \rangle_x$ denotes the spatial average and $\beta_1$ and $\beta_2$ are condition-specific parameters that also depend on timescale $\tau$. Similar power law relationships have been observed with the timescale $\tau$, resulting in relationships of the form
\begin{linenomath*}
\begin{equation}
\langle \dot{\varepsilon}_{\text{tot},\ell\tau}\rangle_t\approx \alpha_1(\ell) \tau^{-\alpha_2(\ell)}, \label{eq:temporal_scaling}
\end{equation}
\end{linenomath*}
for coefficients $\alpha_1$ and $\alpha_2$ that depend on spatial scale $\ell$. Importantly, the value of $\beta_2$ is a quantitative measure of heterogeneity in the deformation field.  Similarly, $\alpha_2$ is a measure of intermittency. As detailed in \citet{Girard2009}, model predictions should have deformation fields that exhibit this power law behavior and have coefficients $\beta_2$ and $\alpha_2$ within realistic bounds.   Notice that $\langle \dot{\varepsilon}_{\text{tot},\ell\tau}\rangle_x$ still depends on time and $\langle \dot{\varepsilon}_{\text{tot},\ell\tau}\rangle_t$ still depends on space; we therefore compute $\langle \dot{\varepsilon}_{\text{tot},\ell\tau}\rangle_x$ at multiple times and $\langle \dot{\varepsilon}_{\text{tot},\ell\tau}\rangle_t$ at multiple locations.

\subsection{Numerical Approximation.}
In practice, we do not have access to a continuous strain rate field $\dot{\varepsilon}(x,t)$ because of limited observations and model discretizations.  To enable computation, we therefore need to approximate both the strain rate tensor $\dot{\varepsilon}(x,t)$ itself and subsequently the integral in \eqref{eq:avg_sr}.  \citet{Girard2010} employs what amounts to a piecewise constant approximation of $\dot{\varepsilon}(x,t)$ on a regular model grid and then approximates \eqref{eq:avg_sr} over a square domain $\mathcal{X}_{\ell}(x) = [x_1-\ell/2, x_1+\ell_2]\times[x_2-\ell/2, x_2+\ell/2]$ by finding cells with centroids in $\mathcal{X}_{\ell}(x)$ and then computing an area-weighted average of the strain rates in these fields.\footnote{The authors of \citet{Girard2010} actually compute an average of the spatial gradient of the velocity field, but because the relationship between velocity gradient and strain rate is linear, this is equivalent to averaging the strain rate.} Because the area of the cells will in general not be $\ell^2$ exactly, the square root of the summed cell areas is used as the ``observed'' length scale $\hat{\ell}$ when computing the power law parameters.  Another approach based on Delaunay triangulations of ``tracer points'' is used for representing $\dot{\varepsilon}(x,t)$ and for approximating \eqref{eq:avg_sr} in \citet{Oikkonen2017} and \citet{Rampal2019}.  Again, the strain rate is piecewise constant, but over triangles in the Delaunay triangulation.  In that work, the averaging window $\mathcal{X}_\ell(x,t)$ is implicitly defined by subsampling the tracer points and creating triangulations with larger cells.   We employ a similar triangular representation of the strain rate tensor but use an explicit spatial average of the strain rate tensor more akin to \citet{Girard2010}.

A DEM simulation gives the position and velocity of each particle at a finite number of times.  For the spatial scaling analysis, we use $\tau=0$ and use the instantaneous particle velocities to compute the strain rate without evaluating the temporal integral in \eqref{eq:avg_sr}.  To approximate $\dot{\varepsilon}(x,t)$, we construct a Delaunay triangulation of the particle centroids, which gives us a triangular mesh with particle velocities corresponding to nodal velocities in this mesh.  As in \cite{Oikkonen2017}, we remove cells in the Delaunay triangulation with a minimum angle of less than $15^\circ$, which could result in poor strain rate approximations and are typically found between particles that are not in contact (i.e., over open water). We also ignore cells based on boundary particles, which do not move in our simulations. Using the nodal velocities, we can then compute cell-wise strain rate tensors using standard finite element machinery (see e.g., \citet{dolfin2010}).  

Let $x^{(i)}$ denote the centroid of cell $i$ in the triangular mesh.  To compute the total deformation rate $\dot{\varepsilon}_{\text{tot},\ell\tau}$ at this point, we use use a circular subdomain $\mathcal{X}_\ell(x^{(i)}) = B_\ell(x^{(i)})$ to define the spatial average, as opposed to the square subdomain employed in \citet{Girard2010}.  The circular subdomain allows us to use KD trees for efficient neighborhood searches. We find all cells in the mesh with centroids $x^{(j)}\in B_\ell(x^{(i)})$ and compute the cell area-weighted average of the strain rates in these cells.  More specifically, 
\begin{linenomath*}
\begin{equation}
    \bar{\varepsilon}_{\ell\tau}^{(i)} = \frac{1}{A_{tot}^{(i)}} \sum_{\{j : x^{(j)}\in B_\ell(x^{(i)})\}} A^{(j)} \dot{\varepsilon}^{(j)},
\end{equation}
\end{linenomath*}
where $A^{(j)}$ is the area of triangle $j$ in the Delaunay triangulation and $A_{tot}^{(i)} \sum A^{(j)}$ is the sum of cell areas for cells intersecting $B_\ell(x^{(i)})$. The length scale associated with this total deformation is given by $\hat{\ell}^{(i)} = \sqrt{A_{tot}^{(i)}}$.  
From $\bar{\varepsilon}_{\ell\tau}^{(i)}$, we can then compute the total deformation rate $\dot{\varepsilon}_{\text{tot},\ell\tau}$ using \eqref{eq:total_def}.

For any length scale $\ell$ and time $t$, we obtain pairs $(\hat{\ell}^{(i)},\langle \dot{\varepsilon}^{(i)}\rangle_{\ell\tau})$ for each cell in the Delaunay triangulation. We use the average of these pairs (over all cells) as an estimate of $\langle \dot{\varepsilon}_{\text{tot},\ell\tau}\rangle_x$ in \eqref{eq:spatial_scaling}.  The process is repeated for multiple length scales ($\ell\in\{5,10,15,20,30,50,80\}$ for our synthetic results) and a least squares fit in log-log space is used to obtain the coefficients $\beta_1$ and $\beta_2$ in the power law.  

The temporal scaling analysis is simpler because the integral over time in \eqref{eq:avg_sr} can be estimated as 
\begin{linenomath*}
\begin{equation}
    \frac{1}{\tau}\int_t^{t+\tau} \dot{\varepsilon}(x,t)dt \approx \frac{1}{2\tau}\left[\nabla(p(x,t+\tau)-p(x,t)) +\nabla(p(x,t+\tau)-p(x,t))^T  \right],
    \label{eq:temporal_integral2}
\end{equation}
\end{linenomath*}
where $p(x,t)$ is a continuous displacement field that we estimate by treating the particle positions as nodal values with piecewise linear finite elements.
We assume that the length scale $\ell=0$, so we can look at cell-wise deformations and do not need to include the spatial averaging in our temporal scaling analysis. To compute the average strain rates, we construct a mesh using the positions at time $t$, then use the change in particle positions to define nodal values for $p(x,t+\tau)-p(x,t)$ and again use standard finite element machinery to compute piecewise constant strain rate tensors in each cell of the mesh (i.e., the right hand side of \eqref{eq:temporal_integral2}). For any cell, the same least squares approach described above for computing $\beta_1$ and $\beta_2$ can then be used to compute the temporal power law parameters $\alpha_1$ and $\alpha_2$ for $\langle \dot{\varepsilon}_{\text{tot},\ell\tau}\rangle_t$ in that cell.  
}

\end{document}